\documentclass[fleqn,usenatbib]{mnras}

\usepackage{newtxtext,newtxmath}
\usepackage[T1]{fontenc}

\DeclareRobustCommand{\VAN}[3]{#2}
\let\VANthebibliography\thebibliography
\def\thebibliography{\DeclareRobustCommand{\VAN}[3]{##3}\VANthebibliography}

\usepackage{graphicx}	% Including figure files
\usepackage{amsmath}	% Advanced maths commands
\usepackage{tabularx}
\usepackage{multirow}
\title[Characterization of interstellar carbon dust analogues]{Characterization of interstellar carbon dust analogues synthesized by dielectric barrier discharge and evolution after irradiation with 3~MeV H\textsuperscript{+}}

\author[I. C. Gerber et al.]{
Ioana Cristina Gerber,$^{1}$
Ilarion Mihaila,$^{1}$
Valentin Pohoata,$^{2}$\thanks{E-mail: vpohoata@uaic.ro (Valentin Pohoata), claire.pirim@univ-lille.fr (Claire Pirim) and ionut.topala@uaic.ro (Ionut Topala)}
Andrei Sandu,$^{2}$
Catalin Agheorghiesei,$^{2}$ \newauthor
Laurentiu Valentin Soroaga,$^{1}$
Decebal Iancu,$^{3}$
Radu Florin Andrei,$^{3}$
Ion Burducea,$^{3}$
Mihai Straticiuc,$^{3}$ \newauthor
Dumitru Duca,$^{4}$
Dmitrii Egorov,$^{4}$
Yvain Carpentier,$^{4}$
Bertrand Chazallon,$^{4}$
Alessandro Faccinetto,$^{5}$ \newauthor
Nicolas Nuns,$^{6}$
Cristian Focsa,$^{4}$
Claire Pirim,$^{4}$\footnotemark[1]
and Ionut Topala$^{2}$\footnotemark[1]
\\
$^{1}$Integrated Center of Environmental Science Studies in the North-Eastern Development Region (CERNESIM), Alexandru Ioan Cuza University of Iasi,\\ Blvd. Carol I No. 11, Iasi 700506, Romania\\
$^{2}$Iasi Plasma Advanced Research Center (IPARC), Faculty of Physics, Alexandru Ioan Cuza University of Iasi, Blvd. Carol I No. 11, Iasi, 700506, Romania\\
$^{3}$Applied Nuclear Physics Department (DFNA), Horia Hulubei National Institute for R\&D in Physics and Nuclear Engineering (IFIN HH),\\ Reactorului 30, Magurele 77125, Romania\\
$^{4}$Univ. Lille, CNRS, UMR 8523 - PhLAM - Physique des Lasers Atomes et Molécules, F-59000 Lille, France\\
$^{5}$Univ. Lille, CNRS, UMR 8522 -PC2A -Physicochimie des Processus de Combustion et de l’Atmosphère, F-59000 Lille, France\\
$^{6}$UMR 8181—UCCS—Unité de Catalyse et Chimie du Solide, CNRS, Centrale Lille, Université d'Artois, Université de Lille, F-59000 Lille, France
}

\date{Accepted 2025 February 17. Received 2025 February 11; in original form 2024 June 20}

\pubyear{2025}

\begin{document}
\label{firstpage}
\pagerange{\pageref{firstpage}--\pageref{lastpage}}
\maketitle

\begin{abstract}
‘Fluffy’ hydrogenated amorphous carbon (a-C:H) was synthesized using a dielectric barrier discharge plasma, driven by nanosecond high voltage pulses at 1 kHz frequency in a helium-butane mixture. The a-C:H samples were characterized by scanning and transmission electron microscopy, laser-assisted and secondary ion mass spectrometry, and Raman and Fourier-transform infrared spectroscopy. We find that a-C:H samples exhibit infrared absorption features in good agreement with those observed for carbonaceous dust in IRAS 08572+3915 galaxy. We discuss their nano to microscale structure and derive their hydrogen to carbon (H/C) ratios from the results obtained by three distinct experimental characterization techniques. Relying on the average H/C value determined by mass spectrometry and Raman spectroscopy, we can then constrain the absorption strengths values to those best corresponding to our dust analogue, and calculate the H/C ratio from the infrared spectra. Altogether, we find that our dust analogue consists of a dominant hydrogen-rich aliphatic network, with small, isolated, aromatic regions. 
The a-C:H dust analogue was then irradiated with 3~MeV H\textsuperscript{+} and subsequently analyzed \textit{ex situ}. Morphological and chemical changes, including the evolution of H/C,  CH\textsubscript{2}/CH\textsubscript{3}, and sp\textsuperscript{2}/sp\textsuperscript{3} ratios, were observed with increasing proton fluence, indicating dehydrogenation and graphitization. Proton bombardment shifted the initial location of a-C:H in the hydrocarbon ternary phase diagram toward the central region defined by IRAS 08572+3915 observations. The decay of the 3.4~$\muup$m band with proton fluence was used to calculate CH destruction cross-sections, results consistent with a direct effect of cosmic rays on the disappearance of the 3.4~$\muup$m band.
\end{abstract}


%%%%%%%%%%%%%%%%%%%%%%%%%%%%%%%%%%%%%%%%%%%%%%%%%%

%%%%%%%%%%%%%%%%% BODY OF PAPER %%%%%%%%%%%%%%%%%%

\section{Introduction}

Carbon-rich asymptotic giant branch stars can replenish the interstellar medium (ISM) with stellar ejecta taking the form of carbon dust grains. The grains are subjected to many interactions with weathering agents, including ultraviolet photons, hydrogen atoms and protons, and are partially transformed into isolated or core-shell structures made of hydrogenated amorphous carbon (a-C:H) and polycyclic aromatic hydrocarbons (PAH) \citep{Pascoli2000-uf,Pendleton2002-tf,Carpentier2012-xj,Chiar2013-qg,Contreras2013-sr}. Such energetic interactions can thus drive morphological and dimensional modifications in carbon dust grains and ices \citep{Bennett2013-hk}. Heating processes in the ISM (photoelectric, UV radiation or X-rays, cosmic rays, gas-grain chemical reactions) can also induce modifications of atomic bonding types and aromatization of a-C:H materials \citep{Gadallah2013-oj,Martinez2020-oo}. Understanding the mechanisms of carbon dust grain formation in the ISM or in circumstellar regions, along with their subsequent energetic or thermal processing as they evolve within their complex environment, is needed to better interpret observational information and apprehend the underlying physical processes at play \citep{Martinez2020-oo,Santoro2020-ta}. Theoretical and experimental studies designed to support the interpretation of scientific observations of such processes are still the strongest approaches in astrophysics, astrochemistry and astrobiology. In this respect, continuous development of laboratory astrophysics studies is essential, although reproducing all processes occurring in astrophysical environments in the laboratory is deemed difficult \citep{Salama2019-ms,Van_Dishoeck2019-ui,Tielens2022-bp}.

A few decades ago, many laboratory works bolstered the hypothesis that a-C:H was the carbon architecture responsible for the near and mid-infrared absorption bands observed in the ISM, particularly for the 3.4~$\muup$m (aliphatic CH stretching), 6.85~$\muup$m, and 7.27~$\muup$m absorption bands (aliphatic CH bending) \citep{Duley1983-bz}. However, a unique molecular structure for a-C:H has not yet been identified and often depends on the production method. In fact, under the single label a-C:H can be found a large group of carbonaceous solids (polymer-like films, diamond-like carbon, graphite-like films and tetrahedrally bonded hydrogenated amorphous carbon), with additionally a relatively large variation of hydrogen to carbon (H/C) ratio and sp\textsuperscript{1}, sp\textsuperscript{2}, sp\textsuperscript{3} contents \citep{Jones1990-cu,Ristein1998-tu,Schultrich2018-qc}. Infrared signatures may not always be sufficient to unambiguously differentiate the hydrogenated carbonaceous solids synthesized under laboratory conditions, as exemplified by the 3.4~$\muup$m nonspecific absorption feature which is common to most laboratory a-C:H \citep{Shinohara2008-bq,Shinohara2018-io,Manis-Levy2014-lu,Asnaz2022-wd}. In space, the 3.4 $\muup$m absorption feature can also appear nonspecific as it is observed in various environments, such as in the ISM – whether it is in the Milky Way or in other galaxies \citep{Imanishi2002-ul,Pendleton2002-tf,Dartois2007-ul}, in the protoplanetary nebula CRL 618 \citep{Chiar1998-jb}, or in smaller objects, such as interplanetary and cometary dust particles or meteorites \citep{Ehrenfreund1991-lj,Matrajt2005-qq,Sandford2006-gk,Munoz_Caro2008-do,Lantz2015-uq}.

The main techniques utilized in the laboratory to produce interstellar carbon dust analogues, in the form of thin films with variable density or porosity and compact, powder-like, solid particles are of two types: plasma enhanced chemical vapor deposition \citep{Furton1999-mn,Kovacevic2005-lg,Contreras2013-sr,Mate2014-go,Molpeceres2017-mh,Gunay2018-el,Gavilan_Marin2020-al,Sciamma-OBrien2020-bm} and laser ablation \citep{Scott1996-op,Mennella1999-ap,Reynaud2001-uy,Llamas-Jansa2007-ob,Biennier2009-kc,Fulvio2017-yw}, sometimes coupled with gas-phase condensation \citep{Mennella2001-sm,Jager2008-sk,Jager2009-fg,Gadallah2011-yj}. However, many other techniques have been described in the literature, such as combustion flames \citep{Pino2008-al,Carpentier2012-xj}, pyrolysis \citep{Reynaud2001-uy,Biennier2009-kc,Fulvio2017-yw}, mechanochemical milling \citep{Dartois2020-ly}, sublimation and quenching \citep{Pendleton2002-tf}, and UV photolysis \citep{Dartois2005-xt,Gadallah2011-yj}. At the nanoscale, disordered molecular structures are characteristic to most interstellar dust analogues and various values of the aromatic to aliphatic ratio are reported depending on the deposition technique. Operational parameters such as the temperature or energy level, precursor molecules, atoms or targets, and pressure are all key factors involved in the experimental design of carbonaceous interstellar dust analogues syntheses and processing.

\begin{table*} % Table 1
	\centering
	\caption{Typical properties of hydrogenated carbonaceous materials deposited using plasma techniques for the various studies on interstellar matter that exhibits 3.4 $\muup$m, 6.85 $\muup$m and 7.27 $\muup$m absorption bands.}
	\label{table:1}
    \renewcommand{\arraystretch}{1.5}
	\begin{tabularx}{\textwidth}[t]{>{\centering\arraybackslash}m{1.8cm}
                                    >{\centering\arraybackslash}m{1.3 cm} 
                                    >{\centering\arraybackslash}m{1.25 cm} 
                                    >{\centering\arraybackslash}m{1.5 cm} 
                                    >{\centering\arraybackslash}m{2.1 cm} 
                                    >{\centering\arraybackslash}m{1.25 cm} 
                                    >{\centering\arraybackslash}m{1.0 cm} 
                                    >{\centering\arraybackslash}m{1.25 cm} 
                                    >{\centering\arraybackslash}m{2.5 cm}} 
		\hline
		Plasma generation method & Precursor &  Pressure (mbar) & Substrate & Product synth. &Thickness ($\muup$m) &  Density (g/cm\textsuperscript{3}) &  H/C & Reference\\
		\hline 
		DC discharge & methane  & 0.26 & NaCl or fused silica & HAC film & 0.3 & 1.5 & 0.5 & \cite{Furton1999-mn}\\
		Microwave discharge & methane & - & NaCl or CaF\textsubscript{2}  & Quenched Carbonaceous Composite & - & - & - & \cite{Goto2000-qn}\\
		Arc discharge or laser ablation & carbon in hydrogen & 10 & KBr & Hydrogenated carbon grains & 0.05, 0.07 & 1.5 & 0.1 – 0.72 & \cite{Mennella2002-nv,Mennella2003-bj}\\
		   Laser ablation & graphite in helium and hydrogen & 3.3 - 26.7 & CaF\textsubscript{2} or KBr & Soot particles & 0.17 & 1.55 & 0.14 - 0.57 & \cite{Jager2008-sk}\\
	RF Plasma enhanced chemical vapour deposition \& Burning propylene with dioxygen & methane and butadiene \& propylene & 0.01 \& 70 & KBr or KCl \& KBr & a-C:H \& soot & 7 - 10 \& 5 & 1.2 \& 1.8 & 1 \& 0.01 & \cite{Godard2011-su}\\
        Laser ablation & graphite in helium and hydrogen & 4.5 & CaF\textsubscript{2} or KBr & HAC particles & - & - & 0.14 - 0.82 & \cite{Gadallah2012-vt}\\
        ICP RF discharge & methane in helium & 0.3 & Silicon & HAC film & 3 & - & - & \cite{Mate2014-go}\\
	ICP RF discharge & methane in helium & 0.32 & Silicon & a-C:H film & 0.42 - 0.52 & 1.2 & 1 & \cite{Mate2016-bu}\\
	ICP RF discharge & methane in helium & 0.3 & Silicon or ZnSe& HAC film & - & - & 1 & \cite{Molpeceres2017-mh}\\
	RF Plasma enhanced chemical vapour deposition & methane & 0.01 & ZnSe &  a-C:H & <10 & 1.2 & 1 & \cite{Dartois2017-aa}\\
  	ICP RF discharge & methane in helium & 0.3 & Silicon & a-C:H film & 0.38 - 1.54 & 1.1 & 1 & \cite{Pelaez2018-dl}\\
        Pulsed discharge nozzle & acetylene or isoprene in argon & - & Petri dish & - & - & - & 1.32 - 1.67 & \cite{Gunay2018-el}\\
        Dielectric Barrier Discharge & methane, ethane, propane, butane in helium & 800 & Graphite or silicon, quartz, and NaCl & a-C:H ‘fluffy’ dust & 100 - 200 & - & 1 & \cite{Hodoroaba2018-gr}\\
        DC magnetron discharge & graphite in argon, then acetylene exposure & 0.1 & SiO\textsubscript{x} or carbon grids or KBr & - & - & - & - & \cite{Santoro2020-ta}\\
       Dielectric Barrier Discharge & methane & 800 & Graphite & a-C:H ‘fluffy’ dust & 100 - 200 & 0.95 & 0.62 - 0.79 & present work\\
		\hline
	\end{tabularx}
\end{table*}

Plasma-based deposition methods offer some unique advantages in interstellar carbon dust analogues production because of the possibility to balance surface and volume reactions, control the ratio of two-body versus three-body processes and the density of radicals and reactants. Most plasma devices employed for carbon dust deposition are low pressure plasma sources, based on capacitively- or inductively-coupled radio frequency discharges, laser ablation plumes, pulsed discharge nozzles, spark and arc discharges, or magnetron discharges. Recently, we demonstrated that an atmospheric pressure plasma source, i.e. a dielectric barrier discharge (DBD), could also be successfully utilized as a synthesis method for ‘fluffy’ interstellar carbon dust analogues under low temperature conditions \citep{Hodoroaba2018-gr}. Along with the differences in plasma techniques, a variety of hydrocarbons, e.g., alkanes, alkenes, alkynes or PAHs, can also be utilized as precursors to carbon dust synthesis (see for example Table 1 in \cite{Contreras2013-sr}). This diversity contributes to synthetizing hydrogenated carbonaceous materials exhibiting distinct properties. Table~\ref{table:1} gathers examples of plasma syntheses and further shows the relationship between plasma operating parameters, the nature of the precursor, and the subsequent properties of the a-C:H materials produced. One parameter characterizing interstellar carbon dust analogues that can be compared to observational data is the H/C ratio. For interstellar dust, this ratio can be assessed from the observed infrared absorption spectra of luminous mid-infrared galaxies, such as IRAS 08572+3915, the latter exhibiting one of the strongest known 3.4~$\muup$m absorption features \citep{Wright1996-dx,Mason2004-jo, Imanishi2006-yc}. Previous H/C estimates for IRAS 08572+3915 were found in the 0.29 to 0.69 range \citep{Dartois2007-ul}. Nevertheless, these H/C ratios should be treated with caution due to the uncertainty in absolute infrared band strength values used in the calculations \citep{Herrero2022-sg}. Table 1 shows that H/C ratios for most interstellar carbon dust analogues fall in the same range or close to the range assessed for IRAS 08572+3915. One can observe a trend for increasing H/C ratio towards values equal to 1 and a trend for decreasing density towards values close to 1~g/cm$\textsuperscript{3}$. Note that in the laboratory, complementary analytical techniques can be used to evaluate H/C ratios, which can help constrain the uncertainty of absolute band strength values within such complex hydrogenated carbonaceous materials.

Evolution of carbon dust in various energetic astrophysical environments (i.e. under exposure to vacuum ultraviolet photons or cosmic rays) was in the spotlight of many experimental studies. Some were devoted to the calculation of destruction cross sections of the 3.4~$\muup$m band by means of energetic processing using (V)UV photons \citep{Mennella2001-al,Alata2014-fy}, ions \citep{Mennella2003-bj,Godard2011-su}, or electrons \citep{Mate2016-bu}. The most recent results of \cite{Godard2011-su} and \cite{Mate2016-bu} suggest that cosmic rays cannot be solely responsible for the disappearance of the 3.4~$\muup$m band (marker of aliphatic C-H bonds) in dense clouds \citep{Herrero2022-sg}.

In this paper we thoroughly characterize the physico-chemical properties of interstellar dust analogues produced in low temperature DBD fed with 15\% butane. We focus our analysis on our analogues’ microscopic features, molecular structure and carbon hybridization, using complementary analytical techniques addressing the particles’ surface and/or bulk composition. To do so, we utilized electron and optical microscopy, two different techniques of mass spectrometry, Raman and Fourier transform infrared (FTIR) spectroscopies. FTIR and electron microscopy were also used to characterize the interstellar dust analogues after their exposure to 3~MeV H\textsuperscript{+} irradiation experiments performed to simulate energetic processing. The analyses are discussed in an astrophysical context, with an emphasis on carbon structure and destruction rates of aliphatic C-H bonds and their relevance to dust evolution in diffuse and dense ISM.

\section{Experimental methods}
\textbf{\textit{Dielectric barrier discharge (DBD) deposition:}} the hydrogenated carbonaceous ‘fluffy’ dust was produced employing DBD to generate a pulsed plasma in helium (grade 4.6) and 15\% butane (grade 2.5), with a total initial pressure of 600~Torr in a closed stainless-steel chamber. Flexible graphite (GoodFellow Co.) was used as substrate in the form of 0.5~mm × 30~mm rectangular strips, homogeneously distributed over the ground electrode. A total of 40 substrate strips were used during each deposition experiment. Further information relative to plasma generation, diagnosis, and monitoring methods are given in a previous work \citep{Hodoroaba2018-gr}. High voltage pulses (1~kHz, 6~kV, 500~ns, positive polarity) were used to periodically drive ionization of the gas mixture between the planar electrodes covered with glass. The available discharge gap between the dielectric layers was 5~mm. Plasma filaments were occasionally visible, sustained by the enhanced local electric field around sharp substrate edges. The average gas temperature was close to room temperature and high energy plasma ions were not major contributors to surface or volume processes. During plasma pulses, electron impact collisions and Penning ionization processes led to the generation of hydrocarbon radicals, which further triggered rich volume and surface chemistries between plasma pulses. As a result, after 6 hours of total deposition time, a black ‘fluffy’ dust was accumulated on all graphite strips, with an average weight gain of 0.6~mg per strip. Macroscopic monitoring of a-C:H growth during the deposition was performed with a camera (Optika C-B5). Dust analogues samples were transferred immediately after deposition to an acrylic storage box equipped with indicating silica gel to minimize exposure to moisture. All characterization techniques were performed \textit{ex situ} on pristine samples duplicates. In addition, the apparent density ($\rho$) of the DBD-produced carbon dust samples was evaluated using the buoyancy method in non-polar solvents.

\textbf{\textit{Microscopy:}} a detailed analysis of the DBD-produced carbon dust morphology was performed with a Quanta FEI 250 scanning electron microscope (SEM). No metal coating was applied to preserve sample morphology. The conductive properties of the DBD-produced carbon dust allowed image acquisition with the electron gun operating at voltages up to 30~kV with no noticeable local charging or image artefacts. In contrast, irradiated samples (after 3~MeV H\textsuperscript{+} exposure) did exhibit slight local charging. For this reason, general surveys of the irradiated samples were performed at 1.5~kV and only higher magnification images were acquired at 30~kV.

In addition, high resolution microscopy was carried out with a FEI Titan electron microscope. The microscope was operated in the scanning transmission electron microscope (STEM) mode, the signal was collected from the backscattered electrons achieving sub-\AA\ resolution. The DBD-produced carbon dust was gingerly scraped off its substrate and transferred to a copper grid (holey carbon film). STEM images were then recorded at different magnifications.

A 4K high accuracy digital microscope (Keyence VHX 6000) was additionally utilized to reconstruct sample surface topography from multi-height focal planes recombination (real time depth composition).

\textbf{\textit{Two-step laser mass spectrometry (L2MS) and time of flight secondary ion mass spectroscopy (TOF-SIMS):}} the molecular content of the DBD-produced carbon dust was studied using a high-resolution two-step laser mass spectrometer (HR-L2MS, Fasmatech S\&T), combining ion cooling, radio frequency (RF) guiding, and a reflectron orthogonal time of flight (Re-oTOF) analyzer, with a maximum mass resolution of about $m/\triangle m \approx 15 000$ at $m/z =  200$. The sample, placed under vacuum ($10^{-8 }$~mbar residual pressure), was irradiated normally by a frequency doubled Nd:YAG laser beam (Quantel Brilliant EaZy, $\lambda$ = 532~nm, 4~ns pulse duration, $\sim$ 120~mJ~cm\textsuperscript{-2} fluence). The instrument was used in ablation mode and thus the desorbed molecules stemmed from both the sample surface and sample bulk. The desorbed compounds formed a gas plume expanding into the vacuum normally to the sample surface, and were ionized by an orthogonal UV laser beam (Quantel Q-smart 850, $\lambda$ = 266~nm, 5~ns pulse duration, 300~mJ~cm\textsuperscript{-2} fluence). The generated ions were then RF-guided to a He collision cell for thermalization and subsequently mass analyzed. A total of 10 mass spectra was acquired in positive polarity across the sample surface. The mass spectra were calibrated using peaks corresponding known hydrogenated species (C\textsubscript{3}H\textsubscript{3}, C\textsubscript{4}H\textsubscript{2}, C\textsubscript{5}H, C\textsubscript{6}H\textsubscript{4} and C\textsubscript{7}H\textsubscript{7}) to provide confident assignments. Only peaks higher than three times the standard deviation of the background noise were included in the data reduction.
Complementary analyses were performed using a ToF-SIMS 5 instrument from ION-TOF GmbH. The DBD-produced carbon dust samples were sputtered with Bi\textsubscript{3}\textsuperscript{+} ions and the generated secondary ions accelerated and analyzed with a time-of-flight mass spectrometer with maximum mass resolution $m/\triangle m \approx 10 000$. The estimated ion dose of $10^{11}$~ions~cm\textsuperscript{-2} was below the threshold of ToF-SIMS static mode, which means that only the uttermost surface layer was analyzed. Mass spectra in positive polarity were recorded at 50~scans/acquisition on a 500~×~500~$\muup$m$^2$ surface with an image resolution of 128~pixels~×~128~pixels. Acquisitions were performed on three-to-five different regions of interest (ROIs) on each sample. The collected mass spectra were aligned, calibrated and normalized by the total relevant ion count. Only peaks higher than three times the standard deviation of the background noise were included in the data reduction. 
Mass defect analysis was performed on mass spectra (for both HR-L2MS and TOF-SIMS) to assign a molecular formula to the recorded accurate mass \citep{Sleno2012-ri,Duca2019-cs}. When mass defect is plotted against nominal mass, species that line up contain a repeating unit, which can simplify the visualization and the interpretation of complex sets of mass spectra (for instance, aliphatic and aromatic hydrocarbons are aligned on different slopes). The global H/C ratio can be calculated from the identified ions in the mass spectra: 
\begin{equation} %Eq:1
    \frac{H}{C}=\frac{N_H}{N_C},\ \ N_X=\sum_iN_{X,i}w_i,\ \ \sum_i w_i=1,\ \ X=H,C
	\label{eq:1}
\end{equation} 
where \textit{N\textsubscript{X,i}} is the number of atoms \textit{X} and \textit{w\textsubscript{i}} the normalized intensity for the corresponding molecular formula \citep{Dobbins1998-gq}. The caveats surrounding this method are: (i) identified ions are the only ones included in the calculation, regardless of the mass spectrometry method utilized, (ii) it is known to slightly overestimate the H/C ratio in TOF-SIMS spectra \citep{Faccinetto2020-om}, and (iii) the relative ionization efficiency is not taken into account in either methods.

\textbf{\textit{Fourier Transformed Infrared (FTIR) Spectroscopy:}} analyses were performed on 2 instruments using 3 different modes to access complementary information. On the first instrument (Bruker Vertex 70), micro-FTIR analyses were carried out on specific DBD-produced carbon dust surface locations in order to reveal the vibrational signatures corresponding to the uppermost molecular sample layers. The spectra were recorded in transmission–reflection mode between 4000~cm\textsuperscript{-1} and 550~cm\textsuperscript{-1} with a resolution of 4~cm\textsuperscript{-1} using the Bruker Vertex 70 FTIR spectrometer coupled to a Hyperion 1000 microscope equipped with $15\times$ (N.A. 0.4) Cassegrain and $4\times$ (N.A. 0.1) objectives. The FTIR spectrometer includes a KBr/Ge beam splitter and the microscope a liquid N\textsubscript{2}-cooled narrow band HgCdTe photoconductor detector. Bare gold mirror backgrounds were recorded from 4000~cm\textsuperscript{-1} to 550~cm\textsuperscript{-1} at atmospheric pressure prior to sample analysis and were used as references in processing sample spectra. The IR beam diameter size was adjusted to 10~$\muup$m using blades mounted within the microscope and focus was adjusted optically to target the sample surface. The same FTIR instrument was later utilized in transmission mode, whereby the DBD-produced carbon dust samples were this time scraped off their substrate and transformed into KBr pellets and then placed within the instrument’s transmission cell. This second mode, where the infrared beam (1~cm beam diameter) passes through the whole pellet, was used to probe sample bulk, in contrast to the previous micro-FTIR analysis where surface features were preponderant. Spectra were recorded as transmittance and then were converted into absorbance.

Further analyses of the DBD-produced carbon dust samples were performed using a second instrument (Jasco FT/IR-4700) equipped with an attenuated total reflectance Fourier transform infrared (ATR-FTIR) accessory (ATR-Pro One). In this configuration, the sample strip was directly placed face down onto a 2.5~mm-diameter germanium crystal so that the IR interacts with the carbon dust at a 45$^\circ$ incident angle to the crystal/sample interface. Sample contact with the crystal was maintained using a dedicated screw on the sample mount. Average spectra of all measurements will be discussed in the results section. While ATR-FTIR spectra were recorded as transmittance, they were subsequently converted into absorbance in selected spectral ranges for the calculation of CH\textsubscript{2}/CH\textsubscript{3}, sp\textsuperscript{2}/sp\textsuperscript{3}, and H/C ratios, and further converted to optical depth values for visualization and comparison purposes with astronomical data from IRAS 08572+3915.

Processing of ATR spectra was further performed to retrieve quantitative information. Specifically, a two-point method was used for baseline removal, followed by a multiple peak fitting procedure using Gaussian components. CH\textsubscript{2}/CH\textsubscript{3}, sp\textsuperscript{2}/sp\textsuperscript{3}, and H/C ratios were calculated using the integrated absorbance of the relevant spectral features \citep{Chiar2013-qg,Molpeceres2017-mh}, ignoring any diamond network and small aromatic domains. However, several corrections have first to be applied. The first correction stems from the ATR method itself, in which the penetration depth ( $p_d$) of evanescent waves into the samples varies with the wavelength of incident infrared radiations. Using an average refractive index value in the infrared range of interest equal to 2.0 for a-C:H materials \citep{Jones2012-zo}, one can estimate an average penetration depth of $\sim$300~nm and $\sim$600~nm for the 3.4~$\muup$m and 6.0~$\muup$m bands, respectively. Accordingly, a reduction factor is used to account for the different volumes sampled at different wavelengths and their consequence on band areas.

A second correction needs to be considered when calculating CH\textsubscript{2}/CH\textsubscript{3} and sp\textsuperscript{2}/sp\textsuperscript{3} ratios. In fact, band areas (A\textsubscript{band}) corresponding to the CH\textsubscript{3} groups, CH\textsubscript{2} groups, aromatic CH groups and olefinic C=C groups must be normalized by the vibrational modes’ absorption strength values \citep{Dartois2007-ul,Chiar2013-qg}. Band areas corresponding to  A\textsubscript{C=C} and the sum (A\textsubscript{CH2} + A\textsubscript{CH3}) were used for calculation of sp\textsuperscript{2} and sp\textsuperscript{3} fractions. Choosing absorption strength values best corresponding to a specific dust analogue is key because the formers are influenced by the density and local electric field, the temperature, local molecular structure, impurities, and average chain length, which makes them known only for a limited number of aliphatic or aromatic molecular structures. Accordingly, only few experimental works were devoted to assessing the absorption strength values at 3.4~$\muup$m for interstellar carbon dust analogues. This was done for instance for interstellar carbon dust analogues produced by laser ablation of graphite \citep{Duley1998-kn} and by a pulsed discharge nozzle method using isoprene and acetylene as precursor gases \citep{Gunay2018-el}. The values determined for these materials were found to be less than half those calculated for small molecules, suggesting that a misuse of absorption strengths describing solids produced in different conditions might lead to deviations from actual CH\textsubscript{2}/CH\textsubscript{3} and sp\textsuperscript{2}/sp\textsuperscript{3} values. It is clear from these experiments that absorption strengths are to be chosen wisely, either after in situ determination within the same experimental conditions, from literature data obtained in very similar conditions, or when H/C ratios can be assessed by complementary experiments or simulated by theoretical models \citep{Molpeceres2017-mh}. Our strategy is then to use H/C ratios determined from complementary spectroscopic approaches (e.g. FTIR and Raman) and ultra-high sensitivity mass spectrometry techniques (e.g. HR-L2MS and TOF-SIMS) to deduce  the absorption strength values best adapted to our DBD-produced carbon dust sample among the values avalaible in the literature \citep{Dartois2007-ul,Chiar2013-qg}. This procedure will ultimately allow us to calculate representative CH\textsubscript{2}/CH\textsubscript{3} and sp\textsuperscript{2}/sp\textsuperscript{3} ratios.

A complementary method to calculate the H/C ratio from FTIR data is to find the positive solution of the following second order polynomial equation, for H/C ratios comprised between 0.3 and 1 \citep{Mennella2002-nv}:
\begin{equation} %Eq:2
    b\left( \frac{H}{C}\right) +\left(a+12b-\frac{1}{m_Hk}\right)\left(\frac{H}{C}\right)+12a=0
	\label{eq:2}
\end{equation}
      
The integrated mass-absorption coefficient for the 3.4~$\muup$m band, i.e. $k$ in equation \ref{eq:2}, is given by the integral of the ratio $K'\left( \tilde{\nu} \right)/ \tilde{\nu}$, with $\tilde{\nu}$ = 2956~cm\textsuperscript{-1} and $m_H$ is the mass of the hydrogen atom. Values of constants in equation \ref{eq:2} are $a = (1.4 \pm 0.1) \times 10^{21}$ cm\textsuperscript{2} and $b = (-1.3 \pm 0.2) \times 10^{21}$ cm\textsuperscript{2} \citep{Mennella2002-nv}. The average mass-absorption coefficient $K'\left( \tilde{\nu} \right)$ was calculated using equation~\ref{eq:3}, where $\rho$ is the apparent density of the sample determined experimentally using the buoyancy method in non-polar solvents, $p_d$ is the penetration depth of evanescent waves into the samples at a given $\tilde{\nu}$, and $\%T$ is the percent transmission:  
\begin{equation} %Eq:3
    K'\left( \tilde{\nu} \right) = \frac{1}{\rho p_d} \ln{\left( \frac{100}{\%T}\right)} 
	\label{eq:3}
\end{equation}

Note that all spectral analyses are performed on a series of 5 to 10 different samples to ensure both the reproducibility of the result, as well as the possibility to identify and exclude any possible outliers (i.e. samples with different material quantities in investigated areas) from the numerical calculations.

\textbf{\textit{Raman spectroscopy:}} analyses were performed with an inVia Reflex spectrometer (Renishaw) equipped with an Olympus microscope (BXFM) \citep{Chazallon2014-rb}. The spectra presented here were obtained by irradiation with a 514.5~nm laser. The laser power was measured at sample and reduced to 21~$\muup$W to avoid thermal sample degradation. Using a 20$\times$ magnification microscope objective (N.A. 0.4), the laser was focused on the sample surface with a spot diameter of $\sim$2~$\muup$m. The spectrometer was calibrated using the Stokes Raman signal of pure Si (520.5~cm\textsuperscript{-1}). Spectra were recorded in extended scan (500-2200~cm\textsuperscript{-1}) and their spectral resolution was $\sim$4~cm\textsuperscript{-1} using the 1800 grooves/mm diffraction grating. A total of 63 spectra were acquired across the DBD-produced dust surfaces. Over 63 spectra, 49 spectra (best signal to noise ratio) served as a basis for H atomic percent calculation. Raman spectra of amorphous carbon materials are commonly characterized by the D (disordered, 1300–1400~cm\textsuperscript{-1}) and G (graphitic, 1550–1650~cm\textsuperscript{-1}) bands \citep{Tuinstra1970-jg,Lespade1984-fn,Wang1990-xp,Ferrari2000-kn,Beyssac2003-bt, Sadezky2005-ii}. The D band emerges in carbon materials exhibiting polyaromatic organic matter with finite-sized in-plane crystallites (i.e., aromatic domain diameter) and/or disrupted crystal symmetry (edges, defects, vacancies) and corresponds to the breathing mode in aromatic rings. The G band characterizes the simultaneous in-plane carbon-carbon stretching mode (E\textsubscript{2g}) in both chains and rings. Both \cite{Casiraghi2005-fy} and \cite{Buijnsters2009-gw} quantified the hydrogen content in hydrogenated amorphous carbon films using an empirical formula involving the photoluminescent background (i.e., slope in $\muup$m, denoted with m) and the maximum intensity of the Raman G band [I(G)], as described in equations~\ref{eq:4} and \ref{eq:5}, respectively, and for hydrogen contents varying approximately between 20\% and 50\%. Both formulas will be used to assess the H atomic percent ($H[at.\%]$) in our samples.
\begin{equation} %Eq:4
    H\left[at.\%\right]_{Casiraghi~05} = 21.7 + 16.6\log\left( \frac{m}{I(G)}\right)
	\label{eq:4}
\end{equation} %Eq: 5
\begin{equation}
    H\left[at.\%\right]_{Buijnsters~09} = 30.0 + 4.6\log\left( \frac{m}{I(G)}\right)
	\label{eq:5}
\end{equation}
\textbf{\textit{Ion irradiation:}} the DBD-produced carbon dust samples were processed on the 3~MV Tandetron\texttrademark\  accelerator system located at the “Horia Hulubei” National Institute for Physics and Nuclear Engineering – IFIN-HH, Magurele, Romania \citep{Burducea2015-cc}. The ion implantation beam line was used to irradiate the DBD-produced carbon dust samples with 3~MeV protons while placed inside a vacuum chamber (10\textsuperscript{-5}~mbar residual pressure). The average beam current, measured with a Faraday cup, was 100~nA. An X–Y beam sweep system was used to raster scan the samples and a four corner Faraday cup assembly was used to define the total scanned area and to measure the implanted ion dose. Carbon tape was used to fix the samples to the wafer target holder, positioned behind the corner Faraday cups. Different fluences, ranging between $1\times10^{14}$ and $1\times10^{16}$~protons~cm\textsuperscript{-2} were used and the sample holder temperature was monitored during the irradiation process. The maximum value of temperature measured was 390~K for $1\times10^{16}$~protons~cm\textsuperscript{-2} fluence, to avoid any thermal processing of interstellar dust analogues. The mean projected range of 3~MeV protons impinging a compact and homogenous carbonaceous material exhibiting the same characteristics as our a-C:H interstellar dust analogues was estimated at around 176~$\muup$m using Stopping and Range of Ions in Matter package (SRIM, \cite{Ziegler2010-un}), which corresponds to a linear energy transfer equal to 19.4~eV~nm\textsuperscript{-1}. The energy deposited per unit mass sample was in the range $2\times10^{13}$~MeV~mg\textsuperscript{-1} to $2\times10^{15}$~MeV~mg\textsuperscript{-1}. After exposure to high energy protons, the DBD-produced carbon dust samples were quickly sealed in small plastic containers, which were themselves placed on an indicating silica gel bed and further sealed in a larger plastic container to minimize air exposure before \textit{ex situ} analyses.

\section{Results}
\subsection{DBD-produced carbon dust description: from the macro to the nanoscale}
Macroscopic monitoring of the DBD-produced carbon dust evidences the formation of conglomerate-like particles free roaming on the substrate surface and slowly increasing in size before settling in what we refer to here as islands. Consequently, the macroscale distribution of the deposited carbon dust is not uniform across the substrate; the material is preferentially found at the extremities and in cluster form, suggesting a 3D carbon island growth. This type of growth can be explained considering the transport of adsorbed seed radicals on 2D atomic arrays, where two limiting cases for molecular materials growth may be envisaged when taking into account the diffusion rate (D) and the deposition flux (F) \citep{Barth2005-fe}: for large D/F ratios, diffusion is favored and the molecular system reaches a minimum energy configuration which can induce a growth regime close to equilibrium conditions, whereas for low D/F values (which apply here), diffusion is prevented and the growth is essentially determined by kinetics, under non-equilibrium conditions. 
The apparent density ($\rho$) of the DBD-produced carbon dust is measured to be about 0.95~g/cm\textsuperscript{3}, which lies at the lower end of density values commonly reported for plasma-produced a-C:H materials (1~-~1.5~g/cm\textsuperscript{3}, details in Table ~\ref{table:1}).

\begin{figure} %Fig1
	\includegraphics[width=0.8\columnwidth]{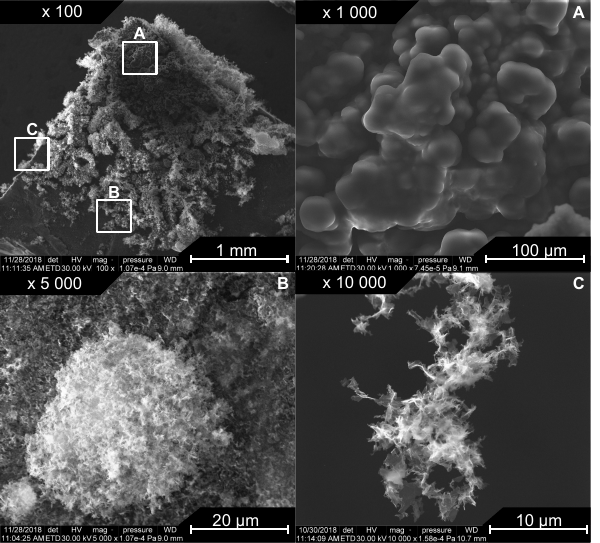}
    \centering
    \caption{SEM micrographs at various magnifications and in various locations on the substrate of the as-deposited DBD-produced carbon dust. Capital letters indicate the regions explored to collect the high magnification images ($\times$1k, $\times$5k, $\times$10k).}
    \label{fig:1}
\end{figure}

Scanning electron microscopy (SEM) reveals two different morphologies at the mesoscale. First, we observe insular structures composed of aggregated a-C:H, their size and abundance increasing towards substrate corners (Fig.~\ref{fig:1}). This observation strongly correlates with the preferential formation and stabilization of discharge streamers in these regions. Second, the material shows a more compact central structure, mainly globular, in the regions where streamers are quite active. This can be attributed to streamer-induced local heating and subsequent transformation of a-C:H conglomerates. Thus, this mesoscopic ordering observed in SEM imaging is strongly influenced by intermolecular interactions and local electric field distribution during deposition.

\begin{figure} %Fig2
	\includegraphics[width=\columnwidth]{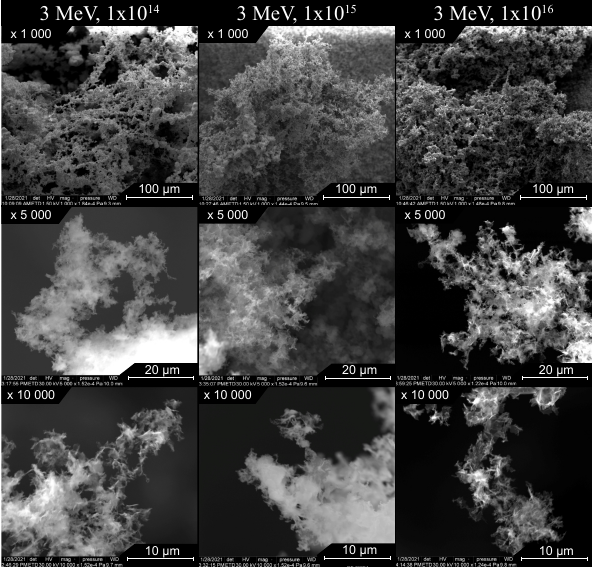}
    \caption{ SEM micrographs of DBD-produced carbon dust samples after 3~MeV H\textsuperscript{+} irradiation with increasing fluences. Column 1 corresponds to $1\times10^{14}$~protons~cm\textsuperscript{-2}, column 2 to $1\times10^{15}$~protons~cm\textsuperscript{-2} and column 3 to $1\times10^{16}$~protons~cm\textsuperscript{-2}.}
    \label{fig:2}
\end{figure}

Figure~\ref{fig:2} shows the morphological differences that appear in SEM micrographs when the DBD-produced carbon dust samples have been subjected to 3~MeV H\textsuperscript{+} with different irradiation fluences. Even at the lowest irradiation dose ($1\times10^{14}$~protons~cm\textsuperscript{-2}), the flakes look damaged and compacted. Also, the flakes now appear etched, the process leaving massive holes that can reach all the way down to the substrate when the flakes are exposed to the highest proton doses. Proton irradiation makes the flakes adhere stronger to each other and gives the previous insular growth a coral-like structure. 

\begin{figure} %Fig3
	\includegraphics[width=0.95\columnwidth]{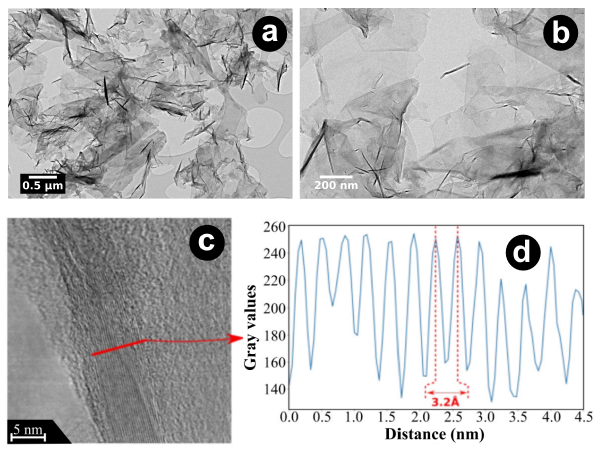}
    \centering
    \caption{A typical crease in STEM images and the corresponding intensity distribution of a cross-section profile in gray level units, shown in red in the left image.}
    \label{fig:3}
\end{figure}

STEM analyses show that some parts of the DBD-produced carbon dust consist of a significant number of "sheets" with numerous creases (dark striae in Figure 3 a and b). Note that the low contrast made only possible to image the edges of graphitic sheets, hereby concealing the amorphous phase. The observed “sheets” are composed of several layers of carbon (10-15 layers), showing a graphite-like structure. The layers can be easily observed in the regions where the sheets are folding and have an approximate thickness of 4-5~nm. The profile of a crease was examined, showing that in Figure~\ref{fig:3} contains 13 carbon layers with an average interplanar distance of 3.2~\AA. This value is in good agreement with values reported in the literature for graphite as well as the one obtained from electron diffraction patterns, i.e. 3.34~\AA\ \citep{Czigany2010-ui}.

\begin{figure} %Fig4
	\includegraphics[width=0.9\columnwidth]{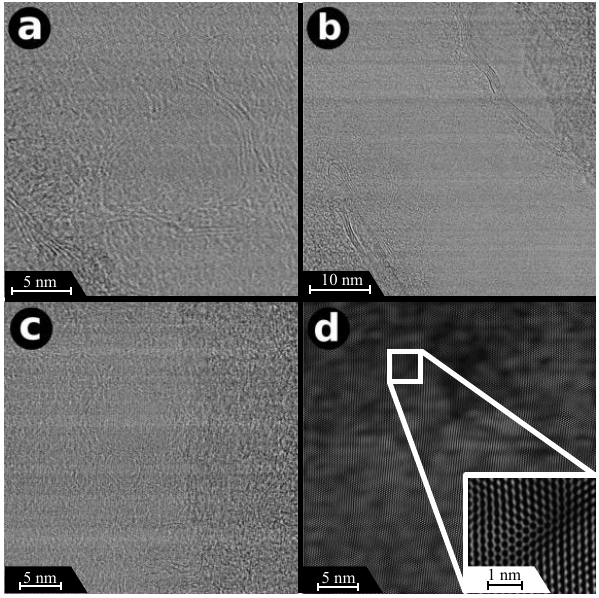}
    \centering
    \caption{High-resolution STEM images (a, b, c) of the interstellar dust analogues and (d) the result of inverse FFT calculated after application of the band pass filter. The inset in (d) is a close-up of the observed patterns.}
    \label{fig:4}
\end{figure}

The graphitic structure of the observed sheets was observed in high-resolution STEM images, Figures \ref{fig:4}a to \ref{fig:4}c. STEM images not only contain information about the lattice in the focal plane of the instrument, but also additional signals originating from out-of-focus layers and amorphous carbon. In order to highlight the structure of graphite sheets within the DBD-produced carbon dust, we subjected the STEM image to an image processing routine in Figure~\ref{fig:4}d. The first processing step was to apply a 2D Fast Fourier Transform (FFT), which revealed 6 high-intensity regions coming from the periodic signal. Then we applied a band-pass filter, and finally, for visualization purposes, we calculated the inverse FFT in order to reconstruct the image associated with the selected frequency range of the filter (Figure~\ref{fig:4}d). This processing effectively removes the noise and leaves only the information corresponding to the periodical lattice. The processed image (Figure~\ref{fig:4}d) clearly shows the honeycomb-like structure of the graphite sheet with the distance between carbon atoms close to the one reported in the literature (140~pm, see the inset in Figure~\ref{fig:4}d). It was shown previously that the high resolution transmission electron microscopy images can be used to identify the degree of crystallinity of carbon materials \citep{Endo1997-hq} and FFT processing was found to be useful in assessing the extent of crystallinity \citep{Lehman2011-oq}. 

\subsection{Determination of H/C ratios}
\begin{figure*} %Fig5
	\includegraphics[width=0.8\textwidth]{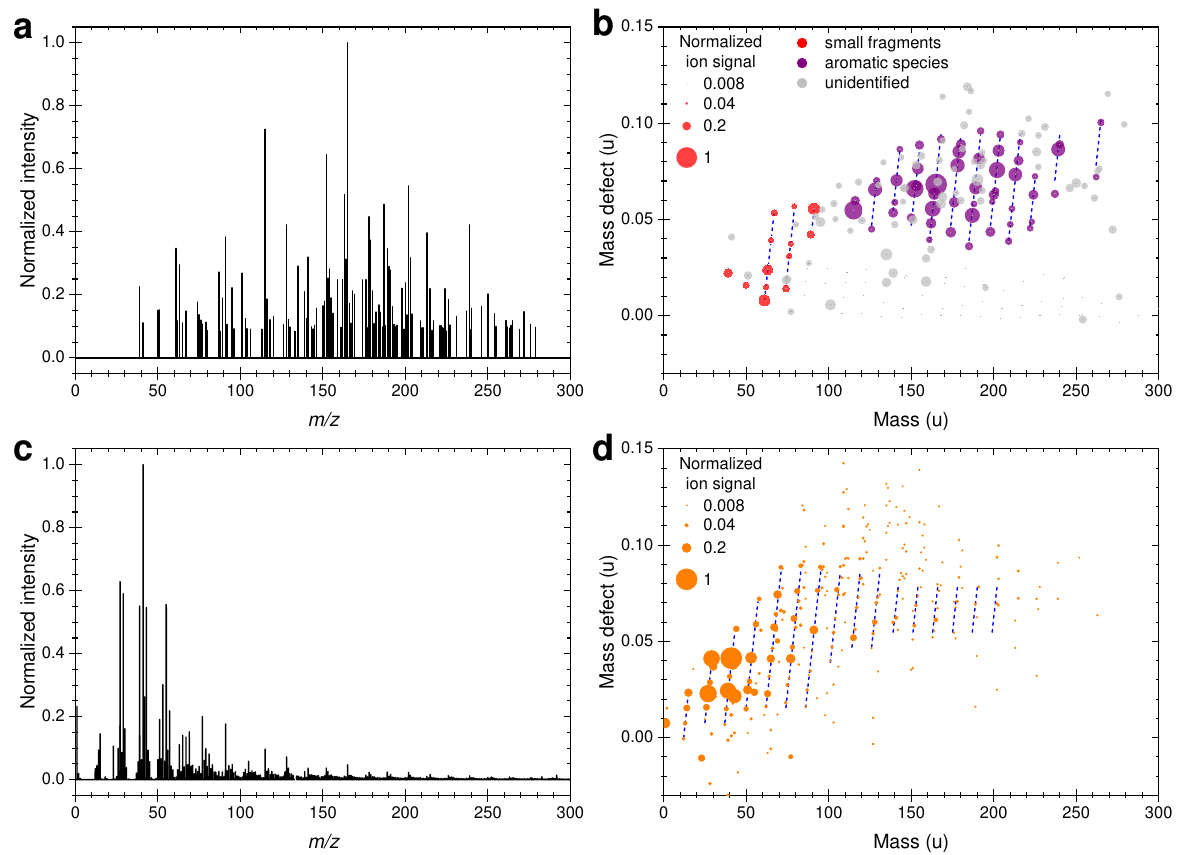}
    \centering
    \caption{As-deposited DBD-produced carbon dust. (a) HR-L2MS normalized synthesis mass spectrum constructed by eliminating species also belonging to the virgin graphite substrate and (b) corresponding mass defects plot. The unidentified species appear in light grey color. (c) ToF-SIMS normalized mass spectrum and (d) corresponding mass defect plot. In the mass defect plots, C\textsubscript{m}H\textsubscript{n}\textsuperscript{+} ions are found aligned on different positive slopes (blue dashed lines). The area of the data points is proportional to the peak intensity.}
    \label{fig:5}
\end{figure*}

The H/C ratio of the DBD-produced carbon dust is obtained from mass spectrometry data. Figure~\ref{fig:5} compares the mass spectra (left column) and the corresponding mass defect plot (right column) obtained independently from HR-L2MS (top row) and ToF-SIMS (bottom row) measurements. 

After mass calibration, 62\% of the detected peak signals are assigned, which corresponds to about 87\% of the total ion count. Peak signals, detected up to around 300 $m/z$, are identified by mass defect analysis following the protocol proposed in \cite{Irimiea2019-xd}. The vast majority of the identified species are assigned to C\textsubscript{m}H\textsubscript{n}\textsuperscript{+} ions that span from small fragments to large polyaromatics, and are highly consistent between the two analyses. No nitrogen bearing species were detected, while only few oxygen containing fragments were identified.

In general, peaks in the mass spectra recur in well-separated groups. The peaks in the same group share the same number of carbon atoms but exhibit a progressively increasing number of hydrogen atoms. Each group is found on a positive slope in the mass defect plot (dashed lines in Fig. \ref{fig:5}b and \ref{fig:5}d). Neighboring groups are separated by gaps that become progressively smaller as $m/z$ increases, and completely disappear above 250 $m/z$.

At low $m/z$, the detected ions include CH\textsubscript{1-3}\textsuperscript{+}, C\textsubscript{2}H\textsubscript{3-5}\textsuperscript{+}, C\textsubscript{3}H\textsubscript{3-7}\textsuperscript{+}, C\textsubscript{4}H\textsubscript{3-8}\textsuperscript{+}, C\textsubscript{5}H\textsubscript{3-9}\textsuperscript{+}, C\textsubscript{6}H\textsubscript{4-7}\textsuperscript{+}, C\textsubscript{7}H\textsubscript{5-7}\textsuperscript{+}, C\textsubscript{8}H\textsubscript{5-7}\textsuperscript{+}  and C\textsubscript{9}H\textsubscript{5-7}\textsuperscript{+}. Here, hydrogen-rich ions are found mixed with hydrogen-poor ions generated from the fragmentation of larger species. At high $m/z$, the most intense peaks in each group are consistent with the patterns expected for PAHs. The main peaks show up along with less intense peaks at [M-1]\textsuperscript{+} or [M-2]\textsuperscript{+} attributed to hydrogen elimination, and to [M+1]\textsuperscript{+} and [M+2]\textsuperscript{+} attributed to the carbon isotopic ions. To each even numbers of carbon atoms corresponds an even $m/z$ value characteristic of benzenoid PAHs (e.g., $m/z$ C\textsubscript{10}H\textsubscript{8}\textsuperscript{+}, C\textsubscript{12}H\textsubscript{8}\textsuperscript{+}, C\textsubscript{14}H\textsubscript{10}\textsuperscript{+}, C\textsubscript{16}H\textsubscript{10}\textsuperscript{+}, C\textsubscript{18}H\textsubscript{10}\textsuperscript{+}, C\textsubscript{20}H\textsubscript{12}\textsuperscript{+} and C\textsubscript{22}H\textsubscript{12}\textsuperscript{+}). In a similar fashion, to each odd carbon numbers corresponds an odd $m/z$ value characteristic of resonance-stabilized cation radicals, markers for instance of PAHs with a 5-member ring (e.g., $m/z$ C\textsubscript{9}H\textsubscript{7}\textsuperscript{+}, C\textsubscript{11}H\textsubscript{9}\textsuperscript{+}, C\textsubscript{13}H\textsubscript{9}\textsuperscript{+}, C\textsubscript{15}H\textsubscript{9}\textsuperscript{+}, C\textsubscript{17}H\textsubscript{11}\textsuperscript{+}, C\textsubscript{19}H\textsubscript{11}\textsuperscript{+}  and C\textsubscript{21}H\textsubscript{11}\textsuperscript{+}).

The H/C ratios are calculated using equation~\ref{eq:1} and are found to be in reasonably good agreement, with 0.62 $\pm$ 0.04  and 0.78 $\pm$ 0.08 from HR-L2MS and ToF-SIMS, respectively. HR-L2MS and TOF-SIMS rely on different desorption and ionization methods. HR-L2MS can desorb molecular species down to several $\muup$m depth, and thus can provide information about both the surface and bulk molecular compositions. In addition, HR-L2MS is optimized for the detection of large polyaromatic hydrocarbons like PAHs that are ionized with a favorable resonant two photon ionization (R2PI) scheme at a wavelength of 266~nm, thereby producing mass spectra featuring reduced fragmentation. In contrast, ToF-SIMS uses high-energy Bi\textsubscript{3}\textsuperscript{+} ions at low dose (static mode) that allow the sputtering of species stemming from the first atomic layers of the sample (1-3~nm). Although the relative ionization efficiencies (RIE) of the various molecular species are not taken into account for H/C calculations in either HR-L2MS or ToF-SIMS, both of them give H/C ratios in reasonable agreements with other methods giving OC/TC ratios \citep{Delhaye2017-gy} or H/C ratios \citep{Dobbins1998-gq, Faccinetto2020-om}. This is made possible by averaging an extended $m/z$ range which tends to cancel out any RIE variations. Figure~\ref{fig:5} shows that HR-L2MS and ToF-SIMS mass spectra feature different distributions of peak intensities. This is due to the different ionization methods, which lead to distinct fragmentation processes. However, equation~\ref{eq:1} shows that the number of atoms of a given element X is conserved for H/C calculations, and thus the total number of atoms remains the same regardless of the fragmentation process. Consequently, the H/C ratios can be seen as independent on the method.

\begin{figure} %Fig6
	\includegraphics[width=\columnwidth]{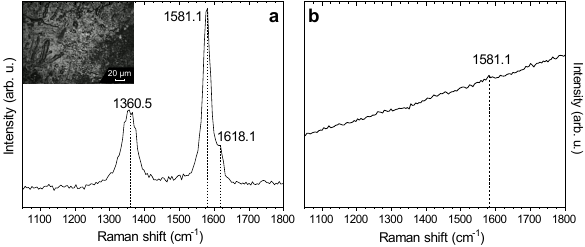}
    \centering
    \caption{Raman spectra acquired across the graphite sheet substrate a) before DBD deposition and b) after DBD deposition in an area optically assessed to be devoid of particle but featuring a DBD-induced coating. Laser power is 21~$\muup$mW in both cases and intensity is in arbitrary units. Inset in Figure~\ref{fig:6}a is an optical image of the pristine graphite sheet substrate taken with the 20$\times$ objective.}
    \label{fig:6}
\end{figure}

\begin{figure} %Fig7
	\includegraphics[width=\columnwidth]{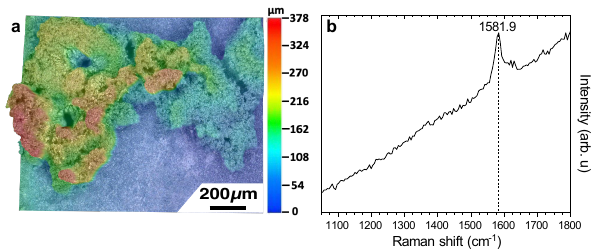}
    \centering
    \caption{DBD Carbon dust a) 3-D surface reconstruction as a color-coded height-map showing its ‘fluffy’ character at the mesoscale and b) corresponding Raman spectrum showing the G-band and a steep background photoluminescence slope.}
    \label{fig:7}
\end{figure}

Raman spectra of the pristine graphite substrate before DBD exposure were first recorded as references and are shown in Figure~\ref{fig:6}. The Raman spectrum of graphite is characterized by the D (disordered, 1300–1400 cm~\textsuperscript{-1}) and G (graphitic, 1550–1650~cm\textsuperscript{-1}) bands \citep{Tuinstra1970-jg,Lespade1984-fn,Wang1990-xp,Ferrari2000-kn,Beyssac2003-bt, Sadezky2005-ii}. Parts of the substrate devoid of carbon dust after DBD exposure feature now a light surface coating of brownish color at near-grazing light incidence. The corresponding Raman spectra exhibit a steep fluorescence background almost obscuring graphite’s G-band Raman signature (Figure~\ref{fig:6}b). Such fluorescence background may originate from covalently bond clusters of PAHs \citep{Ferrari2001-pt} possibly formed via surface hydrocarbon radicals’ chemistry or may be triggered by the insertion of heteroatoms in the carbon lattice \citep{Pal2015-yo}. The previous HR-L2MS and ToF-SIMS analyses indicated the presence of heteroatoms, but only at trace levels. Further analyses of carbon particles discussed below will help shed some light on the fluorescence origin. The DBD-produced carbon dust was first observed through a digital microscope to reconstruct the sample’s surface topography after multi-height focal planes recombination (Figure~\ref{fig:7}a). The topography shows ‘fluffy’ aggregated particles at the macroscale and corroborates previous SEM observations. The 3-D height profile shows that when focused on the particles no contribution from the substrate is expected in Raman spectra, since the sample height (>~$\sim$ 100~$\muup$m) exceeds the theoretical depth of analysis ($\sim$ 5~$\muup$m, at 514.5~nm, using a 20$\times$ objective with a numerical aperture of 0.4, and a refractive index of 2.0 for the carbon dust \citep{Jones2012-zo}).

\begin{figure}
	\includegraphics[width=0.9\columnwidth]{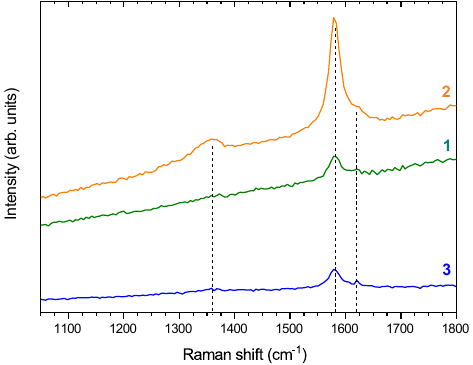}
    \centering
    \caption{ Sequence of 3 Raman spectra displaying the same DBD carbon dust spot subjected to different laser powers. Spectrum 1 was acquired with a laser power of 21~$\muup$W and is followed by spectrum 2 showing the same area exposed afterwards to 121~$\muup$W. Spectrum 3 is obtained exposing the same area again to the lowest laser power of 21~$\muup$W. Spectra 1 and 3 are directly comparable and illustrate the effect of laser power on fluorescence.}
    \label{fig:8}
\end{figure}

Akin to what was observed on graphite after DBD exposure, Raman spectra from the DBD-produced carbon dust exhibit fluorescence (Figure~\ref{fig:7}b). This suggests the existence of aromatics such as condensable PAHs, whose presence was detected in both HR-L2MS and TOF-SIMS. Further studies have been carried out to evaluate their sensibility to thermal degradation or photobleaching. This was done by increasing the  laser fluence in a stepwise manner when analyzing the same spot. Specifically, once selected, a particle spot was subjected to consecutive Raman excitation laser exposures and Raman spectra, all acquired at the same lowest laser power, were subsequently compared (Figure~\ref{fig:8}). This experiment shows that an increase in laser power from 21~$\muup$W to 121~$\muup$W was sufficient to kill most of the fluorescence while preserving Raman information, hereby suggesting a fragile or volatile character of the species generating the observed photoluminescence. In addition, these observations support a surface occurrence, i.e., the presence of these aromatic fluorophores within the uppermost molecular layers of the DBD-produced carbon dust, as confirmed by mass spectrometry. 

Most of the time, the fluorescence is not strong enough to cover Raman signatures, but the ‘fluffy’ nature of the particles can challenge getting proper Raman signal because some sampling spots are off laser focus. In fact, the small laser beam spot diameter ($\sim$2~$\muup$m) and the relatively shallow Raman sampling depth make the resulting Raman spectrum quite sensitive to local roughness. To circumvent this obstacle and obtain meaningful statistics, 63 spectra were acquired across the particles’ surfaces. Hydrogen atomic percents were calculated using equations~\ref{eq:4} and \ref{eq:5} and are shown in table 1. While the first value lies outside the 20-50 H [at.\%] range used to obtain the analytical form of equation~\ref{eq:4} \citep{Casiraghi2005-fy}, one might expect the linearity to remain valid nearby the interval \citep{Adamopoulos2004-oi}. In addition, if one were to make the assumption that carbon atoms total up the atomic percent to 100, then one can derive an average H/C ratio (Table~\ref{table:2}). The H/C ratio of $0.67\pm0.01$ is more in line with the ratios earlier derived from mass spectrometry analyses (Table~\ref{table:2}).

\begin{table}%Table 2
	\centering
	\caption{Atomic percent of hydrogen H [at.\%] and H/C ratio determined from Raman, TOF-SIMS, and HR-L2MS analyses.}
	\label{table:2}
    \renewcommand{\arraystretch}{1.3}
	\begin{tabularx}{\columnwidth}{lcc} 
		\hline
		\ & H [at.\%] & H/C\\
		\hline
		Raman - equation~\ref{eq:4} \citep{Casiraghi2005-fy} & $57.9\pm2.5$ & $1.38\pm0.06$\\
		Raman - equation~\ref{eq:5} \citep{Buijnsters2009-gw} & $40.0\pm0.7$ & $0.67\pm0.01$\\
		TOF-SIMS & - & $0.78\pm0.08$\\
        HR-L2MS & - & $0.62\pm0.04$\\
        \hline
	\end{tabularx}
\end{table}

\begin{figure}
	\includegraphics[width=0.9\columnwidth]{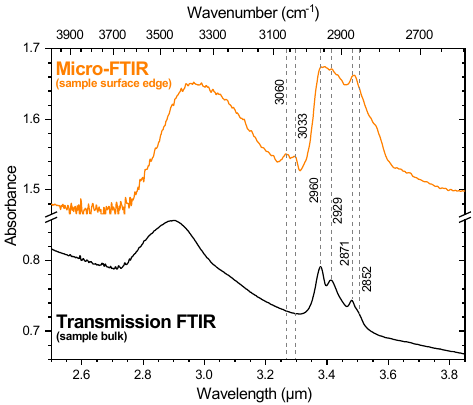}
    \centering
    \caption{FTIR spectra acquired in micro-FTIR mode (top, orange) and in macro FTIR transmission mode with, for the latter, the sample embedded within a KBr pellet (bottom, black). The aromatic C-H stretches (3000-3100~cm\textsuperscript{-1}) emerge only in the micro-FTIR spectrum. The aliphatic CH\textsubscript{x} stretches are all present in both spectra: CH\textsubscript{3} asymmetric ($\sim$2960~cm\textsuperscript{-1}), CH\textsubscript{2} asymmetric ($\sim$2929~cm\textsuperscript{-1}), CH\textsubscript{3} symmetric ($\sim$2871~cm\textsuperscript{-1}), CH\textsubscript{2} symmetric ($\sim$2852~cm\textsuperscript{-1}). Interferences from moisture appear as a broad band between 3200-36001~cm\textsuperscript{-1} in both spectra.}
    \label{fig:9}
\end{figure}

FTIR experiments tailored to probe either sample surface (micro-FTIR) or sample bulk (macro FTIR with sample embedded in KBr pellet and analyzed in transmission mode) further showed that the spectral features expected from the vibration of such aromatic structures exhibited weak signatures when analyzed in micro-FTIR but did not show in transmission mode (Figure~\ref{fig:9}). This is illustrated by the presence of the aromatic C-H stretching vibration (between 3000~cm\textsuperscript{-1} and 3100~cm\textsuperscript{-1} in PAHs \citep{Leger1984-jr, Allamandola1985-ku}) which only emerges in micro-FTIR and, of note, mostly on the edges of the ‘fluffy’ islands of the DBD-produced carbon dust (Figure~\ref{fig:9}).

\begin{figure*}
	\includegraphics[width=0.75\textwidth]{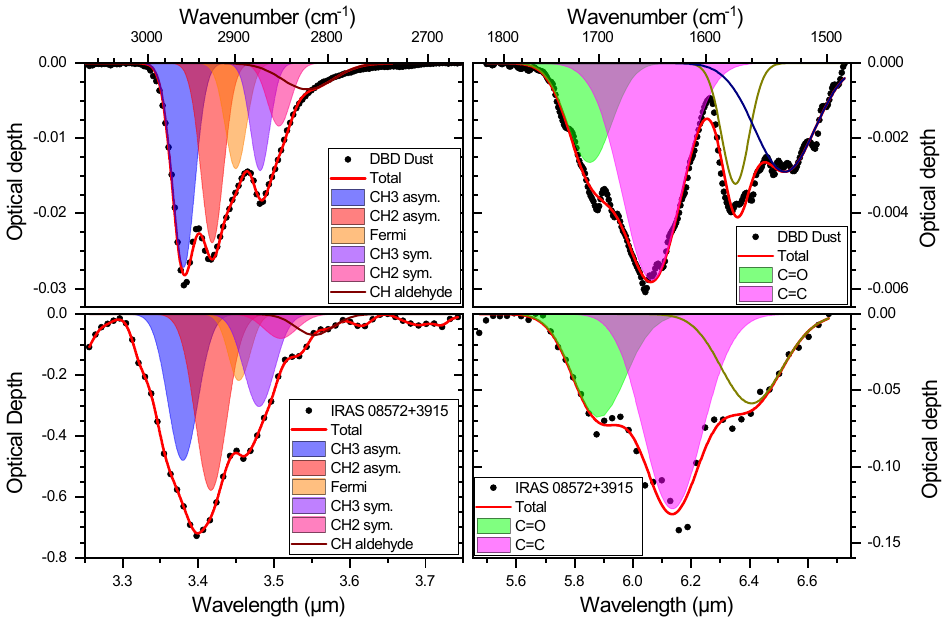}
    \centering
    \caption{Peak fitting of the 3.4~$\muup$m and 6.0~$\muup$m bands in ATR-FTIR spectra. Comparison between the DBD-produced carbon dust sample (top row) and IRAS 08572+3915 (bottom row), the latter taken from \citep{Brown2014-tb}.}
    \label{fig:10}
\end{figure*}

The DBD-produced carbon dust samples exhibit in their ATR-FTIR spectra absorption bands of interest with respect to astrophysical data (Figure~\ref{fig:10}). Among them are the bands emerging between 3000~cm\textsuperscript{-1} and 900~cm\textsuperscript{-1}, with maxima at 2956~cm\textsuperscript{-1} (3.38~$\muup$m), 1455~cm\textsuperscript{-1} (6.87~$\muup$m), and 1375~cm\textsuperscript{-1} (7.27~$\muup$m). Other absorption bands are detailed in our previous works \citep{Hodoroaba2018-gr, Gerber2019-nb}. The ATR-FTIR spectra show no spectral features assigned to vibration modes expected from aromatic structures or sp\textsuperscript{2} hybridized carbon \citep{Ristein1998-tu, Reynaud2001-uy,Dartois2005-xt, Pino2008-al,Carpentier2012-xj,Molpeceres2017-mh}. Specifically, the following bands are not observable: 3000~cm\textsuperscript{-1}~-~3100~cm\textsuperscript{-1} (or 3.3~$\muup$m, aromatic CH stretching), 1560~cm\textsuperscript{-1}~-~1660~cm\textsuperscript{-1} (or 6.2~$\muup$m, aromatic C=C stretching), 1160~cm\textsuperscript{-1} (or 8.6~$\muup$m, aromatic C=C–H in plane bending), 885~cm\textsuperscript{-1}~-~750~cm\textsuperscript{-1} (or 11.3~$\muup$m~-~13.3~$\muup$m solo, duo, trio, and quartet modes of aromatic CH in plane bending). The 3.4~$\muup$m band is assigned to aliphatic -CH stretching modes and is described with the highest accuracy only after a peak fitting procedure is applied, decomposing the signal into five distinct Gaussian components of variable central wavenumbers and full width at half-maximum (FWHM) (Figure~\ref{fig:10}). The corresponding assignments are compiled in Table~\ref{table:3}. 

\begin{table}%Table 3
	\centering
	\caption{Signal decomposition of the DBD-produced carbon dust ATR-FTIR spectrum and corresponding assignments in the 3000~-~2700~cm\textsuperscript{-1} spectral range.}
	\label{table:3}
    \renewcommand{\arraystretch}{1.3}
	\begin{tabular}{ccc} 
		\hline
		Band position (cm\textsuperscript{-1}) & FWHM (cm\textsuperscript{-1}) & Assignment\\
		\hline
		2960 - 2956 & 35 – 19 & asymmetric CH\textsubscript{3} \\
		2935 – 2925 & 34 – 24 & asymmetric CH\textsubscript{2} \\
		2911 – 2898 & 34 – 26 & 2\textsuperscript{nd} harmonic bending CH\textsubscript{2} \\
        2872 – 2868 & 33 – 25 & symmetric CH\textsubscript{3} \\
		2853 – 2842 & 32 – 27 & symmetric CH\textsubscript{2} \\
		\hline
	\end{tabular}
\end{table}

Our spectra are best fitted only including the contribution from the Fermi resonance (2911~cm\textsuperscript{-1}~-~2898~cm\textsuperscript{-1}) which corresponds to the second harmonic of the aliphatic CH\textsubscript{2} scissoring and that is expected for alkyl chains at around 2~$\times$~1455~cm\textsuperscript{-1}. It was previously emphasized that the analysis of spectral features from complex organic materials cannot be simply performed by considering independent vibrations and the analysis of asymmetric CH\textsubscript{2} mode should include a Fermi resonance peak to account for the red wing contribution around 2900~cm\textsuperscript{-1} \citep{Dartois2005-xt,Dartois2007-ul}. The olefinic C=C stretch was found in the 1680~cm\textsuperscript{-1}~-~1648~cm\textsuperscript{-1} range (FWHM~=~121~cm\textsuperscript{-1}~-~58~cm\textsuperscript{-1}), together with a contribution of carbonyl (R\textsubscript{2}C=O) groups, centered at 1710~cm\textsuperscript{-1}. Note that the oxygen content of the interstellar carbon dust analogues probed by mass spectrometry techniques is very low, while in FTIR spectra the C=O stretching mode is present. These carbonyl groups formed from traces in the plasma working gas mixture or during exposure of samples to air, even in very small amounts, are detected in FTIR spectra due to the significant difference in the intrinsic strength of C=O stretches, as compared to C-H stretches \citep{Pendleton2002-tf, Kovacevic2005-lg, Jager2008-sk,Gadallah2012-vt, Carpentier2012-xj,Fulvio2017-yw, Gunay2018-el,Hodoroaba2018-gr}.   
The spectra of this DBD-produced dust are in good agreement with the astronomical observations recorded in the 3.4~$\muup$m and 6.0~$\muup$m band ranges, as shown in Figure~\ref{fig:10}.

From the ATR-FTIR spectra of the DBD-produced carbon dust, we calculated using equation~\ref{eq:2} an H/C ratio of $0.79\pm0.20$. This value is in line with the previous values calculated from the results given by mass spectrometry analyses (considering only the values of H/C~<~1): $0.78\pm0.08$ from ToF-SIMS data, $0.62\pm0.04$ from HR-L2MS data, $0.67\pm0.01$ from Raman spectra using equation~\ref{eq:5}, which gives an average H/C value of $0.72\pm0.20$. Now that the H/C ratio is known, it is possible to choose the absorption strength values from the literature \citep{Dartois2007-ul, Chiar2013-qg} best representing the DBD-produced carbon dust samples. In fact, using the absorption strength values from \cite{Dartois2007-ul} ($12.5\times10^{-18}$~cm~group\textsuperscript{-1} for CH\textsubscript{3} groups, $8.4\times10^{-18}$~cm~group\textsuperscript{-1} for CH\textsubscript{2} groups plus Fermi resonance, $0.4\times10^{-18}$~cm~atoms\textsuperscript{-1}  for C=C groups) and the integrated absorbance of Gaussian components obtained from peak fitting, we obtain H/C~=~$0.72\pm0.04$, consistent with the average H/C ratio from the other techniques. Without this a priori knowledge derived from a multitechnique analysis, the H/C values derived from the ATR-FTIR spectra would range from 0.33 to 1.37. This result shows that a proper selection and cross checks of absorption strength values for relevant carbon-containing chemical groups is a prerequisite before accurate CH\textsubscript{2}/CH\textsubscript{3}, H/C, sp\textsuperscript{2}/sp\textsuperscript{3} ratios calculations can be derived from FTIR laboratory data of astrophysical relevance. The absorption strength values chosen for the DBD-produced carbon dust will serve as a basis for the FTIR study of the evolution of its carbon structure upon 3~MeV H\textsuperscript{+} irradiation.

\section{Evolution after irradiation with 3~MeV H\textsuperscript{+}: astrophysical implications}
\subsection{Carbon structure}
After proton irradiation we observe a decrease of absorbance values in FTIR spectra, together with the modification of relative band ratios (Table~\ref{table:4}). It can be observed that depending on fluence values the sp\textsuperscript{2}/sp\textsuperscript{3} ratio increases, this being usually attributed to a graphitization process, while the H/C ratio decreases due to dehydrogenation by proton bombardment.

\begin{table*}%Table 4
	\centering
	\caption{Evolution of integrated absorbance, integrated mass absorption coefficient of the 3.4~$\muup$m band $k$, and CH\textsubscript{2}/CH\textsubscript{3}, H/C, sp\textsuperscript{2}/sp\textsuperscript{3} ratios, after 3~MeV H\textsuperscript{+} irradiation of the DBD-produced carbon dust samples. A 10\% uncertainty is estimated for calculated CH\textsubscript{2}/CH\textsubscript{3}, H/C, and sp\textsuperscript{2}/sp\textsuperscript{3} ratios.}
	\label{table:4}
    \renewcommand{\arraystretch}{1.3}
	\begin{tabular}{ccccccccc} 
		\hline
		\multirow{2}{*}{Fluence (protons cm\textsuperscript{-2})} & \multirow{2}{*}{t\textsubscript{irr} (h)} & \multicolumn{3}{c}{Integrated absorbance (cm\textsuperscript{-1})} & \multirow{2}{*}{k (cm\textsuperscript{2}/g} & \multirow{2}{*}{CH\textsubscript{2} / CH\textsubscript{3}} & \multirow{2}{*}{H/C} & \multirow{2}{*}{sp\textsuperscript{2}/sp\textsuperscript{3}} \\
		\cline{3-5}
         &  & CH\textsubscript{3} as. & CH\textsubscript{2} as. + Fermi.res. & C=C &  &  &  &  \\
        \hline
		0 & 0 & 0.89 & 1.18 & 0.19 & 88.46 & 2.0 & 0.72 & 2.23\\
		$1.0\times10^{14}$ & 0.2 & 0.68 & 0.63 & 0.09 & 58.69 & 1.4 & 0.91 & 1.67\\
        $3.0\times10^{14}$ & 0.5 & 0.36 & 0.48 & 0.12 & 43.77 & 2.0 & 0.53 & 3.4\\
		$6.0\times10^{14}$ & 0.9 & 0.36 & 0.5 & 0.09 & 27.80 & 2.1 & 0.68 & 2.43\\
  	$1.0\times10^{15}$ & 1.5 & 0.26 & 0.44 & 0.09 & 40.55 & 2.5 & 0.58 & 2.94\\
        $4.0\times10^{15}$ & 6.1 & 0.19 & 0.3 & 0.07 & 38.73 & 2.3 & 0.50 & 3.61\\
		$1.0\times10^{16}$ & 16.1 & 0.11 & 0.23 & 0.06 & 25.65 & 3.1 & 0.45 & 3.99\\
		\hline
	\end{tabular}
\end{table*}

\begin{figure}
	\includegraphics[width=\columnwidth]{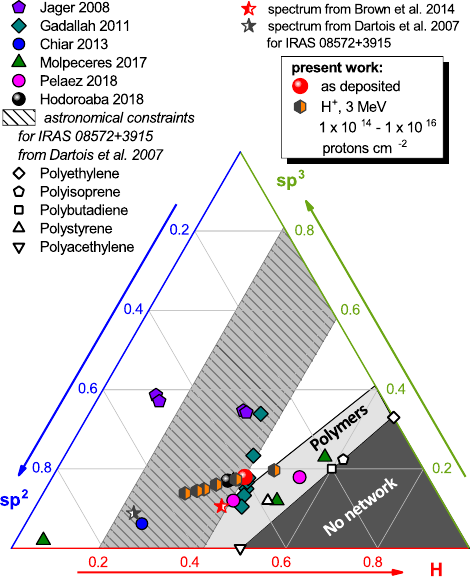}
    \centering
    \caption{ Ternary phase diagrams for materials containing hydrogen (H), C (sp\textsuperscript{2}) and C (sp\textsuperscript{3}), studied as interstellar dust analogues. Source of data points and constraints: \citep{Dartois2007-ul,Brown2014-tb, Jager2008-sk,Gadallah2011-yj,Chiar2013-qg,Molpeceres2017-mh,Pelaez2018-dl,Hodoroaba2018-gr}. }
    \label{fig:11}
\end{figure}

The various forms of carbon-based materials can be described structurally taking into account the hydrogen, sp\textsuperscript{2}, and sp\textsuperscript{3} contents. These three quantities, normalized to 1, are usually plotted as a ternary diagram as shown in Figure~\ref{fig:11}, and also proposed for discussions on the physical properties of interstellar carbon dust analogues \citep{Dartois2007-ul,Chiar2013-qg,Pelaez2018-dl}. Some distinct regions can be clearly identified in the ternary diagram: 1) the “no network” region characterized by a high hydrogen content; 2) the polymers region, characterized by appropriate values of H, sp\textsuperscript{2}, sp\textsuperscript{3} content for polymers to form; 3) the regions nearby sp\textsuperscript{2} and sp\textsuperscript{3} apexes, corresponding to materials close to exhibiting pure graphite and pure diamond structures, respectively. The visual analysis of the ternary diagram and the scattering of various candidates for interstellar dust analogues emphasize the grouping into two families: the polymer-like materials and the hydrogenated amorphous carbons materials. In the specific case of IRAS 08572+3915, the H content between 0.19~–~0.40 derived from observations induces constraints and highlights a specific region of the ternary diagram \citep{Dartois2007-ul}. There is a good agreement with the properties of many laboratory-produced hydrogenated amorphous carbon dust analogues, with structural models based on dominant aliphatic network, with small, isolated, aromatic regions or analogues based on polyaromatic networks, with significant aliphatic content as endings. One can note the data clustering in the ternary diagram from many reported interstellar dust analogues at the intersection of the polymers area with that of IRAS 08572+3915.

It is possible to lock the DBD-produced carbon dust as a-C:H exhibiting a branched aliphatic structure with contributions from multiple C-C bonds. Further processing, using various energetic particles or radiation, usually leads to dehydrogenation and graphitization of interstellar dust analogues. Under the conditions studied in this work, the 3~MeV H\textsuperscript{+} irradiation of the DBD-produced carbon dust induces limited changes in H and sp\textsuperscript{2} contents but a marked left shift in the location of the processed dust inside the region delimited by the constraints from IRAS 08572+3915 astronomical observations.

\subsection{Destruction rates of aliphatic CH bonds}
\begin{figure}
	\includegraphics[width=\columnwidth]{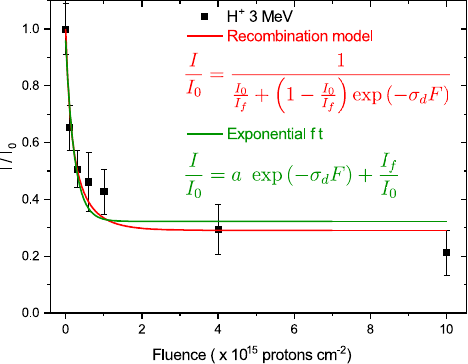}
    \centering
    \caption{\textit{Ex situ} analysis of the 3.4~$\muup$m band area decay after 3~MeV proton bombardment, using a phenomenological exponential decay and the hydrogen recombination model.}
    \label{fig:12}
\end{figure}

After irradiation with 3~MeV protons and \textit{ex situ} analyses, we have observed a gradual decrease of the 3.4~$\muup$m band integrated area as a function of proton fluence with a saturation tendency for high fluences (Figure~\ref{fig:12}). The behavior is qualitatively similar to that described in other in situ energetic processing of a-C:H samples using 30~keV He\textsuperscript{+} \citep{Mennella2003-bj}, 0.2~MeV and 10~MeV H\textsuperscript{+} , 50~MeV C\textsuperscript{5+}, 91~MeV C\textsuperscript{6+}, 85~MeV Si\textsuperscript{7+}, 100~MeV Ni\textsuperscript{9+}, 160~MeV I\textsuperscript{21+} \citep{Godard2011-su} or 5 keV electrons \citep{Mate2016-bu}. The plot of 3.4~$\muup$m band intensity decay versus protons fluence is presented in Figure~\ref{fig:12}. The use of previously proposed data analysis methods, i.e. the exponential fit and the hydrogen recombination model \citep{Godard2011-su, Mate2016-bu} returns the following values for destruction cross section ($\sigma_d$): $1.69\times10^{-15}$~cm\textsuperscript{2} from the recombination model and $4.40\times10^{-15}$~cm\textsuperscript{2} from the exponential fit. The value of asymptotic relative band intensity (I\textsubscript{0}/I\textsubscript{f}) is close to 0.3 using both data fitting methods, leading to a value of 43~\AA\textsuperscript{3} for the characteristic recombination volume within the solid, value typical for ion and electron irradiation experiments \citep{Mate2016-bu,Godard2011-su}. It is worth to stress that the values of destruction cross sections were obtained in \textit{ex situ} studies on ‘fluffy’ a-C:H interstellar dust analogues, experimental data of this type being unavailable in literature.

Ion bombardment was carried out at near room temperature. The macroscopic thickness of ‘fluffy’ samples exposed to irradiation ranges between 100 and 200~$\muup$m \citep{Hodoroaba2018-gr} . Although this is comparable to the simulated ion penetration depth of 176~$\muup$m for a compact material with similar density and stoichiometry, an accurate determination of the samples' thickness is not possible due to their morphology. While ion processing can generally be considered uniform, it is possible that certain regions of the sample may not receive enough radiation due to variations in morphology. The reduction of the 3.4~$\muup$m band intensity after ion irradiation was up to 70\%, reaching a saturation point beyond which no further reduction occurs. \cite{Mennella2003-bj} reported a similar finding at 300 K and a lower sample temperature during ion processing ( i.e. 12 K) was leading to a further reduction of up to 85\%, while complete reduction was not achieved.

The characteristic times for the destruction of aliphatic CH bonds, $\tau_{d,CR}$, under the action of cosmic rays can be calculated as the inverse of the cosmic rays destruction rate, $R_{d,CR}$. This last physical quantity might be described, using a monoenergetic (1~MeV) proton beam \citep{Mennella2003-bj}, as the product of destruction cross section $\sigma_d$(1~MeV) and the effective flux of protons $\phi_p(1MeV)$. Using the hypothesis of direct proportionality between the destruction cross section ($\sigma_d$) and the stopping cross section ($S_p$), we can write: 

\begin{equation}
    \tau_{d,CR}=\left[ \frac{S_p(1MeV)}{S_p(3MeV)}\sigma_d(3MeV)\phi_p(1MeV)\right]^{-1}
	\label{eq:6}
\end{equation}

\begin{table*} %Table 5
	\centering
	\caption{Comparison of available data for destruction rates by cosmic rays for aliphatic CH bonds and destruction time in diffuse regions or dense clouds.}
	\label{table:5}
    \renewcommand{\arraystretch}{1.3}
	\begin{tabular}{ccccc}
    \hline
    \multirow{2}{3cm}{Aliphatic CH bonds destruction by cosmic rays} & \multirow{2}{3 cm}{Aliphatic CH bonds destruction rates (yr\textsuperscript{-1})} & \multicolumn{2}{c}{Destruction time (Myr)} & \multirow{2}{*}{Reference} \\ \cline{3-4}
                      &                   & Diffuse Regions & Dense Clouds &                   \\  \hline
    \multirow{2}{*}{Unlikely} &  $(0.95 - 10.4)\times10^{-9}$ & 100 - 1000  & 100 - 1000 & \cite{Godard2011-su} \\
                      & $(0.59 - 2.08)\times10^{-9}$ & 200 - 500 & 300 - 900 &  \cite{Mate2016-bu} \\  \hline
    \multirow{3}{*}{Probable} & $2.52\times10^{-8}$ & 40 & & \ \ \cite{Dartois2017-aa} \\
                            &   $0.54\times10^{-7}$ & 20 & 30 & \cite{Mennella2003-bj} \\
                            & $(0.53 - 1.39)\times10^{-7}$ & 4 - 10 & 7 - 20 & present work \\ \hline
\end{tabular}
\end{table*}

Using the values of destruction cross section ($\sigma_d$), from both the recombination model and the exponential fit method, the values of characteristic times for destruction of aliphatic CH bonds are in the range 1~-~15~millions of years (Myr), i.e., within the typical lifetime of diffuse regions (100~Myr) and dense clouds (30~Myr) \citep{Jones1994-io}. The following numerical values were used for calculations, returned by SRIM code \citep{Ziegler2010-un} applied for physical characteristics of our interstellar dust analogues ($\rho_{a-C:H}=0.95$~g~cm\textsuperscript{-3}, H/C~=~1): S\textsubscript{p}(3~MeV)~=~0.119~MeV~mg\textsuperscript{-1}~cm\textsuperscript{2}, S\textsubscript{p}(1~MeV)~=~0.259~MeV~mg\textsuperscript{-1}~cm\textsuperscript{2}. The analysis of available data from previous studies (i.e. Table 5 in \cite{Mate2016-bu} and Section 4.1 in \cite{Dartois2017-aa}) is allowing to group the results into two families (Table~\ref{table:5}): data that support the scenario of unlikely aliphatic CH bonds destruction by cosmic rays \citep{Godard2011-su, Mate2016-bu}, and data that support a probable aliphatic CH bonds destruction by cosmic rays (present work, \cite{Dartois2017-aa}, \cite{Mennella2003-bj}). The relationship between the CH destruction cross-section ($\sigma_d$) and the stopping cross-section ($S_p$) was previously investigated (i.e. Figure 9 in \cite{Dartois2017-aa}). However, establishing a direct correlation between the available $\sigma_d$ data at high and low $S_p$ values remains challenging. Additionally, \cite{Dartois2017-aa} suggested that smaller particles may have a larger destruction cross-section than predicted by the aforementioned proposed correlation. The analogue synthesized and tested in this study, generically named ‘fluffy’ dust analogue, consists at nanoscale of an agglomeration of numerous individual flakes (Figure 3). The calculated $\sigma_d$ is an order of magnitude larger than the value predicted by the power law used by \cite{Dartois2017-aa} to fit the ion cross section data. One possible explanation for this discrepancy is the morphological characteristics of the analogue used in this study. In this context, the present study improves the existing literature by providing new data in the low $S_p$ range, thereby improving the overall understanding of this dependence.

\section{Conclusions}

Using a dielectric barrier discharge in He/C\textsubscript{4}H\textsubscript{10} mixtures we produced ‘fluffy’ a-C:H samples exhibiting features in the infrared absorption spectra (e.g., the 3.~$\muup$m band) in good agreement with those of carbonaceous dust from IRAS 08572+3915. These interstellar dust analogues were characterized using a variety of microscopy, mass spectrometry and vibrational spectroscopy techniques in order to gain some insights into their structure, from the macro to the nano scale, and derive their H/C ratio from multiple experimental methods. The average H/C value served then as reference to choose the absorption strength values that allowed the retrieval of the closest matching H/C value from FTIR spectra. In fact, calculation methods based on band intensities measurements of known absorption strengths were debated recently as they can return results with significant deviations if the absorption strengths available in the literature are not adapted to the studied solid dust analogue. We show here that a thorough multitechnique analysis from which can be derived H/C ratios can support the choice of absorption strengths values best adapted to the studied carbon dust analogue. Upon irradiation with 3 MeV H\textsuperscript{+}, the DBD-produced carbon dust shows morphological and chemical changes. The microscale structure appears modified and indications of ion etching are visible on all investigated samples. The H/C, CH\textsubscript{2}CH\textsubscript{3}, and sp\textsuperscript{2}/sp\textsuperscript{3} ratios show an evolution with the proton fluence, whereby the H/C ratio tend to decrease due to the proton bombardment-induced dehydrogenation, whereas the sp\textsuperscript{2}/sp\textsuperscript{3} ratio increases upon graphitization of the sample. The location of the DBD-produced carbon dust in the ternary phase diagram falls close to many other reported data for dust analogues and is shown to shift further in the IRAS 08572+3915 constrained region upon proton bombardment. Based on the \textit{ex situ} study of the DBD-produced carbon dust, we were able to measure the intensity decay of the 3.4~$\muup$m band as a function of proton fluence and calculate the CH destruction cross sections. Comparison with the lifetimes of diffuse and dense regions in astrophysical environments showed that the observed intensity decay is in good agreement with the basic hydrogen recombination model. Finally, the results are discussed in the light of the relevant literature. 

\section*{Acknowledgements}

Ion beam experiments measurements have been performed at 3~MV Tandetron\texttrademark\ accelerator from “Horia Hulubei” National Institute for Physics and Nuclear Engineering (IFIN-HH) and were supported by the Romanian Government Programme through the National Programme for Installations of National Interest (IOSIN) and all three authors from IFIN-HH (M. S., D. I. and R. A.) were supported from the Nucleus programme (PN 19 06 02 01 and PN 19 06 02 02). 
We acknowledge the use of FEI TITAN Themis for TEM images at Plateforme de Microscopie Électronique de Lille (PMEL) of Université de Lille, hosted by Institut Michel-Eugène Chevreul (CNRS FR2638). The authors also acknowledge the CaPPA project (Chemical and Physical Properties of the Atmosphere) funded by the French National Research Agency (ANR) through the PIA (Programme d’Investissement d’Avenir) under contract $\ll$ANR-11-LABX-0005-01$\gg$, the Ministry of Higher Education and Research, Hauts de France Regional Council, and European Regional Development Fund (ERDF) through the Contrat de Projets Etat Region (CPER CLIMIBIO) for their financial support. I.T. also acknowledges the CaPPA project and the Faculty of Sciences and Technologies of the university of Lille for their visiting scientist fellowship programs. The authors would also like to acknowledge the contribution of the Centre d' Etudes et de Recherches Lasers et Applications (CERLA) platform for the materials, equipment and support staff. 

\section*{Data Availability}
The data underlying this article will be shared on reasonable request to the corresponding authors.
 
%Inclusion of a Data Availability Statement is a requirement for articles published in MNRAS. Data Availability Statements provide a standardised format for readers to understand the availability of data underlying the research results described in the article. The statement may refer to original data generated in the course of the study or to third-party data analysed in the article. The statement should describe and provide means of access, where possible, by linking to the data or providing the required accession numbers for the relevant databases or DOIs.



\bibliographystyle{mnras}
\bibliography{MN-24-1349-MJ} % if your bibtex file is called example.bib

\begin{thebibliography}{}
\makeatletter
\relax
\def\mn@urlcharsother{\let\do\@makeother \do\$\do\&\do\#\do\^\do\_\do\%\do\~}
\def\mn@doi{\begingroup\mn@urlcharsother \@ifnextchar [ {\mn@doi@}
  {\mn@doi@[]}}
\def\mn@doi@[#1]#2{\def\@tempa{#1}\ifx\@tempa\@empty \href
  {http://dx.doi.org/#2} {doi:#2}\else \href {http://dx.doi.org/#2} {#1}\fi
  \endgroup}
\def\mn@eprint#1#2{\mn@eprint@#1:#2::\@nil}
\def\mn@eprint@arXiv#1{\href {http://arxiv.org/abs/#1} {{\tt arXiv:#1}}}
\def\mn@eprint@dblp#1{\href {http://dblp.uni-trier.de/rec/bibtex/#1.xml}
  {dblp:#1}}
\def\mn@eprint@#1:#2:#3:#4\@nil{\def\@tempa {#1}\def\@tempb {#2}\def\@tempc
  {#3}\ifx \@tempc \@empty \let \@tempc \@tempb \let \@tempb \@tempa \fi \ifx
  \@tempb \@empty \def\@tempb {arXiv}\fi \@ifundefined
  {mn@eprint@\@tempb}{\@tempb:\@tempc}{\expandafter \expandafter \csname
  mn@eprint@\@tempb\endcsname \expandafter{\@tempc}}}

\bibitem[\protect\citeauthoryear{Adamopoulos, Robertson, Morrison  \&
  Godet}{Adamopoulos et~al.}{2004}]{Adamopoulos2004-oi}
Adamopoulos G.,  Robertson J.,  Morrison N.~A.,   Godet C.,  2004, \mn@doi [J.
  Appl. Phys.] {10.1063/1.1811397}, 96, 6348

\bibitem[\protect\citeauthoryear{Alata, Cruz-Diaz, Mu{\~n}oz~Caro  \&
  Dartois}{Alata et~al.}{2014}]{Alata2014-fy}
Alata I.,  Cruz-Diaz G.~A.,  Mu{\~n}oz~Caro G.~M.,   Dartois E.,  2014, \mn@doi
  [Astron. Astrophys. Suppl. Ser.] {10.1051/0004-6361/201323118}, 569, A119

\bibitem[\protect\citeauthoryear{Allamandola, Tielens  \& Barker}{Allamandola
  et~al.}{1985}]{Allamandola1985-ku}
Allamandola L.~J.,  Tielens A. G. G.~M.,   Barker J.~R.,  1985, \mn@doi
  [Astrophys. J.] {10.1086/184435}, 290, L25

\bibitem[\protect\citeauthoryear{Asnaz, Kohlmann, Folger, Greiner  \&
  Benedikt}{Asnaz et~al.}{2022}]{Asnaz2022-wd}
Asnaz O.~H.,  Kohlmann N.,  Folger H.,  Greiner F.,   Benedikt J.,  2022,
  \mn@doi [Plasma Process. Polym.] {10.1002/ppap.202100190}, 19, 2100190

\bibitem[\protect\citeauthoryear{Barth, Costantini  \& Kern}{Barth
  et~al.}{2005}]{Barth2005-fe}
Barth J.~V.,  Costantini G.,   Kern K.,  2005, \mn@doi [Nature]
  {10.1038/nature04166}, 437, 671

\bibitem[\protect\citeauthoryear{Bennett, Pirim  \& Orlando}{Bennett
  et~al.}{2013}]{Bennett2013-hk}
Bennett C.~J.,  Pirim C.,   Orlando T.~M.,  2013, \mn@doi [Chem. Rev.]
  {10.1021/cr400153k}, 113, 9086

\bibitem[\protect\citeauthoryear{Beyssac, Goff{\'e}, Petitet, Froigneux, Moreau
   \& Rouzaud}{Beyssac et~al.}{2003}]{Beyssac2003-bt}
Beyssac O.,  Goff{\'e} B.,  Petitet J.-P.,  Froigneux E.,  Moreau M.,   Rouzaud
  J.-N.,  2003, \mn@doi [Spectrochim. Acta A Mol. Biomol. Spectrosc.]
  {10.1016/s1386-1425(03)00070-2}, 59, 2267

\bibitem[\protect\citeauthoryear{Biennier, Georges, Chandrasekaran, Rowe,
  Bataille, Jayaram, Reddy  \& Arunan}{Biennier et~al.}{2009}]{Biennier2009-kc}
Biennier L.,  Georges R.,  Chandrasekaran V.,  Rowe B.,  Bataille T.,  Jayaram
  V.,  Reddy K. P.~J.,   Arunan E.,  2009, \mn@doi [Carbon N. Y.]
  {10.1016/j.carbon.2009.07.050}, 47, 3295

\bibitem[\protect\citeauthoryear{Brown et~al.,}{Brown
  et~al.}{2014}]{Brown2014-tb}
Brown M. J.~I.,  et~al., 2014, \mn@doi [ApJS] {10.1088/0067-0049/212/2/18},
  212, 18

\bibitem[\protect\citeauthoryear{Buijnsters, Gago, Jim{\'e}nez, Camero,
  Agull{\'o}-Rueda  \& G{\'o}mez-Aleixandre}{Buijnsters
  et~al.}{2009}]{Buijnsters2009-gw}
Buijnsters J.~G.,  Gago R.,  Jim{\'e}nez I.,  Camero M.,  Agull{\'o}-Rueda F.,
   G{\'o}mez-Aleixandre C.,  2009, \mn@doi [J. Appl. Phys.]
  {10.1063/1.3103326}, 105, 093510

\bibitem[\protect\citeauthoryear{Burducea et~al.,}{Burducea
  et~al.}{2015}]{Burducea2015-cc}
Burducea I.,  et~al., 2015, \mn@doi [Nucl. Instrum. Methods Phys. Res. B]
  {10.1016/j.nimb.2015.07.011}, 359, 12

\bibitem[\protect\citeauthoryear{Carpentier et~al.,}{Carpentier
  et~al.}{2012}]{Carpentier2012-xj}
Carpentier Y.,  et~al., 2012, \mn@doi [Astron. Astrophys. Suppl. Ser.]
  {10.1051/0004-6361/201118700}, 548, A40

\bibitem[\protect\citeauthoryear{Casiraghi, Ferrari  \& Robertson}{Casiraghi
  et~al.}{2005}]{Casiraghi2005-fy}
Casiraghi C.,  Ferrari A.~C.,   Robertson J.,  2005, \mn@doi [Phys. Rev. B
  Condens. Matter] {10.1103/PhysRevB.72.085401}, 72, 085401

\bibitem[\protect\citeauthoryear{Chazallon, Ziskind, Carpentier  \&
  Focsa}{Chazallon et~al.}{2014}]{Chazallon2014-rb}
Chazallon B.,  Ziskind M.,  Carpentier Y.,   Focsa C.,  2014, \mn@doi [J. Phys.
  Chem. B] {10.1021/jp507789z}, 118, 13440

\bibitem[\protect\citeauthoryear{Chiar, Pendleton, Geballe  \& Tielens}{Chiar
  et~al.}{1998}]{Chiar1998-jb}
Chiar J.~E.,  Pendleton Y.~J.,  Geballe T.~R.,   Tielens A.~G.,  1998, \mn@doi
  [Astrophys. J.] {10.1086/306318}, 507, 281

\bibitem[\protect\citeauthoryear{Chiar, {A G G}, Adamson  \& Ricca}{Chiar
  et~al.}{2013}]{Chiar2013-qg}
Chiar J.~E.,  {A G G} Adamson A.~J.,   Ricca A.,  2013, \mn@doi [ApJ]
  {10.1088/0004-637X/770/1/78}, 770, 78

\bibitem[\protect\citeauthoryear{Contreras \& Salama}{Contreras \&
  Salama}{2013}]{Contreras2013-sr}
Contreras C.~S.,  Salama F.,  2013, \mn@doi [ApJS] {10.1088/0067-0049/208/1/6},
  208, 6

\bibitem[\protect\citeauthoryear{Czig{\'a}ny \& Hultman}{Czig{\'a}ny \&
  Hultman}{2010}]{Czigany2010-ui}
Czig{\'a}ny Z.,  Hultman L.,  2010, \mn@doi [Ultramicroscopy]
  {10.1016/j.ultramic.2010.02.005}, 110, 815

\bibitem[\protect\citeauthoryear{Dartois, Mu{\~n}oz~Caro, Deboffle, Montagnac
  \& d'Hendecourt}{Dartois et~al.}{2005}]{Dartois2005-xt}
Dartois E.,  Mu{\~n}oz~Caro G.~M.,  Deboffle D.,  Montagnac G.,   d'Hendecourt
  L.,  2005, \mn@doi [Astron. Astrophys. Suppl. Ser.]
  {10.1051/0004-6361:20042094}, 432, 895

\bibitem[\protect\citeauthoryear{Dartois et~al.,}{Dartois
  et~al.}{2007}]{Dartois2007-ul}
Dartois E.,  et~al., 2007, \mn@doi [Astron. Astrophys. Suppl. Ser.]
  {10.1051/0004-6361:20066572}, 463, 635

\bibitem[\protect\citeauthoryear{Dartois, Chabot, Pino, Béroff, Godard,
  Severin, Bender  \& Trautmann}{Dartois et~al.}{2017}]{Dartois2017-aa}
Dartois E.,  Chabot M.,  Pino T.,  Béroff K.,  Godard M.,  Severin D.,  Bender
  M.,   Trautmann C.,  2017, \mn@doi [Astron. Astrophys. Suppl. Ser.]
  {10.1051/0004-6361/201629646}, 599, A130

\bibitem[\protect\citeauthoryear{Dartois, Charon, Engrand, Pino  \&
  Sandt}{Dartois et~al.}{2020}]{Dartois2020-ly}
Dartois E.,  Charon E.,  Engrand C.,  Pino T.,   Sandt C.,  2020, \mn@doi
  [Astron. Astrophys. Suppl. Ser.] {10.1051/0004-6361/202037725}, 637, A82

\bibitem[\protect\citeauthoryear{Delhaye et~al.,}{Delhaye
  et~al.}{2017}]{Delhaye2017-gy}
Delhaye D.,  et~al., 2017, \mn@doi [J. Aerosol Sci.]
  {10.1016/j.jaerosci.2016.11.018}, 105, 48

\bibitem[\protect\citeauthoryear{Dobbins, Fletcher  \& Chang}{Dobbins
  et~al.}{1998}]{Dobbins1998-gq}
Dobbins R.~A.,  Fletcher R.~A.,   Chang H.-C.,  1998, \mn@doi [Combust. Flame]
  {10.1016/S0010-2180(98)00010-8}, 115, 285

\bibitem[\protect\citeauthoryear{Duca et~al.,}{Duca et~al.}{2019}]{Duca2019-cs}
Duca D.,  et~al., 2019, \mn@doi [Faraday Discuss.] {10.1039/c8fd00238j}, 218,
  115

\bibitem[\protect\citeauthoryear{Duley \& Williams}{Duley \&
  Williams}{1983}]{Duley1983-bz}
Duley W.~W.,  Williams D.~A.,  1983, \mn@doi [Mon. Not. R. Astron. Soc.]
  {10.1093/mnras/205.1.67P}, 205, 67P

\bibitem[\protect\citeauthoryear{Duley, Scott, Seahra  \& Dadswell}{Duley
  et~al.}{1998}]{Duley1998-kn}
Duley W.~W.,  Scott A.~D.,  Seahra S.,   Dadswell G.,  1998, \mn@doi [ApJ]
  {10.1086/311548}, 503, L183

\bibitem[\protect\citeauthoryear{Ehrenfreund, Robert, d'Hendecourt  \&
  Behar}{Ehrenfreund et~al.}{1991}]{Ehrenfreund1991-lj}
Ehrenfreund P.,  Robert F.,  d'Hendecourt L.,   Behar F.,  1991, Astron.
  Astrophys., 252, 712

\bibitem[\protect\citeauthoryear{Endo, Takeuchi, Hiraoka, Furuta, Kasai, Sun,
  Kiang  \& Dresselhaus}{Endo et~al.}{1997}]{Endo1997-hq}
Endo M.,  Takeuchi K.,  Hiraoka T.,  Furuta T.,  Kasai T.,  Sun X.,  Kiang
  C.-H.,   Dresselhaus M.~S.,  1997, \mn@doi [J. Phys. Chem. Solids]
  {10.1016/s0022-3697(97)00055-3}, 58, 1707

\bibitem[\protect\citeauthoryear{Faccinetto et~al.,}{Faccinetto
  et~al.}{2020}]{Faccinetto2020-om}
Faccinetto A.,  et~al., 2020, \mn@doi [Communications Chemistry]
  {10.1038/s42004-020-00357-2}, 3, 1

\bibitem[\protect\citeauthoryear{Ferrari \& Robertson}{Ferrari \&
  Robertson}{2000}]{Ferrari2000-kn}
Ferrari A.~C.,  Robertson J.,  2000, \mn@doi [Phys. Rev. B Condens. Matter]
  {10.1103/PhysRevB.61.14095}, 61, 14095

\bibitem[\protect\citeauthoryear{Ferrari \& Robertson}{Ferrari \&
  Robertson}{2001}]{Ferrari2001-pt}
Ferrari A.~C.,  Robertson J.,  2001, \mn@doi [Phys. Rev. B Condens. Matter]
  {10.1103/PhysRevB.64.075414}, 64, 075414

\bibitem[\protect\citeauthoryear{Fulvio, G{\'o}bi, J{\"a}ger, Kereszturi  \&
  Henning}{Fulvio et~al.}{2017}]{Fulvio2017-yw}
Fulvio D.,  G{\'o}bi S.,  J{\"a}ger C.,  Kereszturi {\'A}.,   Henning T.,
  2017, \mn@doi [ApJS] {10.3847/1538-4365/aa9224}, 233, 14

\bibitem[\protect\citeauthoryear{Furton, Laiho  \& Witt}{Furton
  et~al.}{1999}]{Furton1999-mn}
Furton D.~G.,  Laiho J.~W.,   Witt A.~N.,  1999, \mn@doi [ApJ]
  {10.1086/308016}, 526, 752

\bibitem[\protect\citeauthoryear{Gadallah, Mutschke  \& J{\"a}ger}{Gadallah
  et~al.}{2011}]{Gadallah2011-yj}
Gadallah K. A.~K.,  Mutschke H.,   J{\"a}ger C.,  2011, \mn@doi [Astron.
  Astrophys. Suppl. Ser.] {10.1051/0004-6361/201015542}, 528, A56

\bibitem[\protect\citeauthoryear{Gadallah, Mutschke  \& J{\"a}ger}{Gadallah
  et~al.}{2012}]{Gadallah2012-vt}
Gadallah K. A.~K.,  Mutschke H.,   J{\"a}ger C.,  2012, \mn@doi [Astron.
  Astrophys. Suppl. Ser.] {10.1051/0004-6361/201219248}, 544, A107

\bibitem[\protect\citeauthoryear{Gadallah, Mutschke  \& J{\"a}ger}{Gadallah
  et~al.}{2013}]{Gadallah2013-oj}
Gadallah K. A.~K.,  Mutschke H.,   J{\"a}ger C.,  2013, \mn@doi [Astron.
  Astrophys. Suppl. Ser.] {10.1051/0004-6361/201220895}, 554, A12

\bibitem[\protect\citeauthoryear{Gavilan~Marin, Bejaoui, Haggmark, Svadlenak,
  de Vries, Sciamma-O'Brien  \& Salama}{Gavilan~Marin
  et~al.}{2020}]{Gavilan_Marin2020-al}
Gavilan~Marin L.,  Bejaoui S.,  Haggmark M.,  Svadlenak N.,  de Vries M.,
  Sciamma-O'Brien E.,   Salama F.,  2020, \mn@doi [Astrophys. J.]
  {10.3847/1538-4357/ab62b7}, 889, 101

\bibitem[\protect\citeauthoryear{Gerber, Chiper, Pohoata, Mihaila  \&
  Topala}{Gerber et~al.}{2019}]{Gerber2019-nb}
Gerber I.~C.,  Chiper A.,  Pohoata V.,  Mihaila I.,   Topala I.,  2019, \mn@doi
  [Proc. Int. Astron. Union] {10.1017/S174392131900749X}, 15, 237

\bibitem[\protect\citeauthoryear{Godard et~al.,}{Godard
  et~al.}{2011}]{Godard2011-su}
Godard M.,  et~al., 2011, \mn@doi [Astron. Astrophys. Suppl. Ser.]
  {10.1051/0004-6361/201016228}, 529, A146

\bibitem[\protect\citeauthoryear{Goto, Maihara, Terada, Kaito, Kimura  \&
  Wada}{Goto et~al.}{2000}]{Goto2000-qn}
Goto M.,  Maihara T.,  Terada H.,  Kaito C.,  Kimura S.,   Wada S.,  2000,
  \mn@doi [Astron. Astrophys. Suppl. Ser.] {10.1051/aas:2000113}, 141, 149

\bibitem[\protect\citeauthoryear{G{\"u}nay, Schmidt, Burton, Af{\c s}ar,
  Krechkivska, Nauta, Kable  \& Rawal}{G{\"u}nay et~al.}{2018}]{Gunay2018-el}
G{\"u}nay B.,  Schmidt T.~W.,  Burton M.~G.,  Af{\c s}ar M.,  Krechkivska O.,
  Nauta K.,  Kable S.~H.,   Rawal A.,  2018, \mn@doi [Mon. Not. R. Astron.
  Soc.] {10.1093/mnras/sty1582}, 479, 4336

\bibitem[\protect\citeauthoryear{Herrero, Jim{\'e}nez-Redondo, Pel{\'a}ez,
  Mat{\'e}  \& Tanarro}{Herrero et~al.}{2022}]{Herrero2022-sg}
Herrero V.~J.,  Jim{\'e}nez-Redondo M.,  Pel{\'a}ez R.~J.,  Mat{\'e} B.,
  Tanarro I.,  2022, \mn@doi [Front. Astron. Space Sci.]
  {10.3389/fspas.2022.1083288}, 9, 1083288

\bibitem[\protect\citeauthoryear{Hodoroaba, Gerber, Ciubotaru, Mihaila,
  Dobromir, Pohoata  \& Topala}{Hodoroaba et~al.}{2018}]{Hodoroaba2018-gr}
Hodoroaba B.,  Gerber I.~C.,  Ciubotaru D.,  Mihaila I.,  Dobromir M.,  Pohoata
  V.,   Topala I.,  2018, \mn@doi [Mon. Not. R. Astron. Soc.]
  {10.1093/mnras/sty2497}, 481, 2841

\bibitem[\protect\citeauthoryear{Imanishi}{Imanishi}{2002}]{Imanishi2002-ul}
Imanishi M.,  2002, \mn@doi [Mon. Not. R. Astron. Soc.]
  {10.1046/j.1365-8711.2000.03873.x}, 319, 331

\bibitem[\protect\citeauthoryear{Imanishi, Dudley  \& Maloney}{Imanishi
  et~al.}{2006}]{Imanishi2006-yc}
Imanishi M.,  Dudley C.~C.,   Maloney P.~R.,  2006, \mn@doi [Astrophys. J.]
  {10.1086/498391}, 637, 114

\bibitem[\protect\citeauthoryear{Irimiea, Faccinetto, Mercier, Ortega, Nuns,
  Therssen, Desgroux  \& Focsa}{Irimiea et~al.}{2019}]{Irimiea2019-xd}
Irimiea C.,  Faccinetto A.,  Mercier X.,  Ortega I.-K.,  Nuns N.,  Therssen E.,
   Desgroux P.,   Focsa C.,  2019, \mn@doi [Carbon N. Y.]
  {10.1016/j.carbon.2018.12.015}, 144, 815

\bibitem[\protect\citeauthoryear{J{\"a}ger, Mutschke, Henning  \&
  Huisken}{J{\"a}ger et~al.}{2008}]{Jager2008-sk}
J{\"a}ger C.,  Mutschke H.,  Henning T.,   Huisken F.,  2008, \mn@doi [ApJ]
  {10.1086/592729}, 689, 249

\bibitem[\protect\citeauthoryear{J{\"a}ger, Huisken, Mutschke, Llamas~Jansa  \&
  Henning}{J{\"a}ger et~al.}{2009}]{Jager2009-fg}
J{\"a}ger C.,  Huisken F.,  Mutschke H.,  Llamas~Jansa I.,   Henning T.,  2009,
  \mn@doi [ApJ] {10.1088/0004-637X/696/1/706}, 696, 706

\bibitem[\protect\citeauthoryear{Jones}{Jones}{2012}]{Jones2012-zo}
Jones A.~P.,  2012, \mn@doi [Astron. Astrophys. Suppl. Ser.]
  {10.1051/0004-6361/201117624}, 540, A2

\bibitem[\protect\citeauthoryear{Jones, Duley  \& Williams}{Jones
  et~al.}{1990}]{Jones1990-cu}
Jones A.~P.,  Duley W.~W.,   Williams D.~A.,  1990, Q. J. R. Astron. Soc., 31,
  567

\bibitem[\protect\citeauthoryear{Jones, Tielens, Hollenbach  \& McKee}{Jones
  et~al.}{1994}]{Jones1994-io}
Jones A.~P.,  Tielens A. G. G.~M.,  Hollenbach D.~J.,   McKee C.~F.,  1994,
  \mn@doi [Astrophys. J.] {10.1086/174689}, 433, 797

\bibitem[\protect\citeauthoryear{Kova{\v c}evi{\'c}, Stefanovi{\'c}, Berndt,
  Pendleton  \& Winter}{Kova{\v c}evi{\'c} et~al.}{2005}]{Kovacevic2005-lg}
Kova{\v c}evi{\'c} E.,  Stefanovi{\'c} I.,  Berndt J.,  Pendleton Y.~J.,
  Winter J.,  2005, \mn@doi [ApJ] {10.1086/428392}, 623, 242

\bibitem[\protect\citeauthoryear{Lantz et~al.,}{Lantz
  et~al.}{2015}]{Lantz2015-uq}
Lantz C.,  et~al., 2015, \mn@doi [Astron. Astrophys. Suppl. Ser.]
  {10.1051/0004-6361/201425398}, 577, A41

\bibitem[\protect\citeauthoryear{Leger \& Puget}{Leger \&
  Puget}{1984}]{Leger1984-jr}
Leger A.,  Puget J.~L.,  1984, Astron. Astrophys., 137, L5

\bibitem[\protect\citeauthoryear{Lehman, Terrones, Mansfield, Hurst  \&
  Meunier}{Lehman et~al.}{2011}]{Lehman2011-oq}
Lehman J.~H.,  Terrones M.,  Mansfield E.,  Hurst K.~E.,   Meunier V.,  2011,
  \mn@doi [Carbon N. Y.] {10.1016/j.carbon.2011.03.028}, 49, 2581

\bibitem[\protect\citeauthoryear{Lespade, Marchand, Couzi  \& Cruege}{Lespade
  et~al.}{1984}]{Lespade1984-fn}
Lespade P.,  Marchand A.,  Couzi M.,   Cruege F.,  1984, \mn@doi [Carbon N. Y.]
  {10.1016/0008-6223(84)90009-5}, 22, 375

\bibitem[\protect\citeauthoryear{Llamas-Jansa, J{\"a}ger, Mutschke  \&
  Henning}{Llamas-Jansa et~al.}{2007}]{Llamas-Jansa2007-ob}
Llamas-Jansa I.,  J{\"a}ger C.,  Mutschke H.,   Henning T.,  2007, \mn@doi
  [Carbon N. Y.] {10.1016/j.carbon.2007.02.032}, 45, 1542

\bibitem[\protect\citeauthoryear{Manis-Levy, Livneh, Zukerman, Mintz  \&
  Raveh}{Manis-Levy et~al.}{2014}]{Manis-Levy2014-lu}
Manis-Levy H.,  Livneh T.,  Zukerman I.,  Mintz M.~H.,   Raveh A.,  2014,
  \mn@doi [Plasma Sci. Technol] {10.1088/1009-0630/16/10/09}, 16, 954

\bibitem[\protect\citeauthoryear{Mart{\'\i}nez et~al.,}{Mart{\'\i}nez
  et~al.}{2020}]{Martinez2020-oo}
Mart{\'\i}nez L.,  et~al., 2020, \mn@doi [Nat Astron]
  {10.1038/s41550-019-0899-4}, 4, 97

\bibitem[\protect\citeauthoryear{Mason, Wright, Pendleton  \& Adamson}{Mason
  et~al.}{2004}]{Mason2004-jo}
Mason R.~E.,  Wright G.,  Pendleton Y.,   Adamson A.,  2004, \mn@doi
  [Astrophys. J.] {10.1086/423316}, 613, 770

\bibitem[\protect\citeauthoryear{Mat{\'e}, Tanarro, Moreno,
  Jim{\'e}nez-Redondo, Escribano  \& Herrero}{Mat{\'e}
  et~al.}{2014}]{Mate2014-go}
Mat{\'e} B.,  Tanarro I.,  Moreno M.~A.,  Jim{\'e}nez-Redondo M.,  Escribano
  R.,   Herrero V.~J.,  2014, \mn@doi [Faraday Discuss.] {10.1039/C3FD00132F},
  168, 267

\bibitem[\protect\citeauthoryear{Mat{\'e}, Molpeceres, Jim{\'e}nez-Redondo,
  Tanarro  \& Herrero}{Mat{\'e} et~al.}{2016}]{Mate2016-bu}
Mat{\'e} B.,  Molpeceres G.,  Jim{\'e}nez-Redondo M.,  Tanarro I.,   Herrero
  V.~J.,  2016, \mn@doi [Astrophys. J.] {10.3847/0004-637X/831/1/51}, 831, 51

\bibitem[\protect\citeauthoryear{Matrajt, Mu{\~n}oz~Caro, Dartois,
  d'Hendecourt, Deboffle  \& Borg}{Matrajt et~al.}{2005}]{Matrajt2005-qq}
Matrajt G.,  Mu{\~n}oz~Caro G.~M.,  Dartois E.,  d'Hendecourt L.,  Deboffle D.,
    Borg J.,  2005, \mn@doi [Astron. Astrophys.] {10.1051/0004-6361:20041605},
  433, 979

\bibitem[\protect\citeauthoryear{Mennella, Brucato, Colangeli  \&
  Palumbo}{Mennella et~al.}{1999}]{Mennella1999-ap}
Mennella V.,  Brucato J.~R.,  Colangeli L.,   Palumbo P.,  1999, \mn@doi [ApJ]
  {10.1086/312302}, 524, L71

\bibitem[\protect\citeauthoryear{Mennella, Brucato  \& Colangeli}{Mennella
  et~al.}{2001a}]{Mennella2001-sm}
Mennella V.,  Brucato J.~R.,   Colangeli L.,  2001a, \mn@doi [Spectrochim. Acta
  A Mol. Biomol. Spectrosc.] {10.1016/S1386-1425(00)00444-3}, 57, 787

\bibitem[\protect\citeauthoryear{Mennella, Mu{\~n}oz~Caro, Ruiterkamp, Schutte,
  Greenberg, Brucato  \& Colangeli}{Mennella et~al.}{2001b}]{Mennella2001-al}
Mennella V.,  Mu{\~n}oz~Caro G.~M.,  Ruiterkamp R.,  Schutte W.~A.,  Greenberg
  J.~M.,  Brucato J.~R.,   Colangeli L.,  2001b, \mn@doi [Astron. Astrophys.
  Suppl. Ser.] {10.1051/0004-6361:20000340}, 367, 355

\bibitem[\protect\citeauthoryear{Mennella, Brucato, Colangeli  \&
  Palumbo}{Mennella et~al.}{2002}]{Mennella2002-nv}
Mennella V.,  Brucato J.~R.,  Colangeli L.,   Palumbo P.,  2002, \mn@doi [ApJ]
  {10.1086/339229}, 569, 531

\bibitem[\protect\citeauthoryear{Mennella, Baratta, Esposito, Ferini  \&
  Pendleton}{Mennella et~al.}{2003}]{Mennella2003-bj}
Mennella V.,  Baratta G.~A.,  Esposito A.,  Ferini G.,   Pendleton Y.~J.,
  2003, \mn@doi [Astrophys. J.] {10.1086/368342}, 587, 727

\bibitem[\protect\citeauthoryear{Molpeceres, Tim{\'o}n, Jim{\'e}nez-Redondo,
  Escribano, Mat{\'e}, Tanarro  \& Herrero}{Molpeceres
  et~al.}{2017}]{Molpeceres2017-mh}
Molpeceres G.,  Tim{\'o}n V.,  Jim{\'e}nez-Redondo M.,  Escribano R.,  Mat{\'e}
  B.,  Tanarro I.,   Herrero V.~J.,  2017, \mn@doi [Phys. Chem. Chem. Phys.]
  {10.1039/c6cp06043a}, 19, 1352

\bibitem[\protect\citeauthoryear{Mu{\~n}oz~Caro, Dartois  \&
  Nakamura-Messenger}{Mu{\~n}oz~Caro et~al.}{2008}]{Munoz_Caro2008-do}
Mu{\~n}oz~Caro G.~M.,  Dartois E.,   Nakamura-Messenger K.,  2008, \mn@doi
  [Astron. Astrophys.] {10.1051/0004-6361:20078879}, 485, 743

\bibitem[\protect\citeauthoryear{Pal}{Pal}{2015}]{Pal2015-yo}
Pal S.~K.,  2015, \mn@doi [Carbon N. Y.] {10.1016/j.carbon.2015.02.035}, 88, 86

\bibitem[\protect\citeauthoryear{Pascoli \& Polleux}{Pascoli \&
  Polleux}{2000}]{Pascoli2000-uf}
Pascoli G.,  Polleux A.,  2000, Astron. Astrophys., 359, 799

\bibitem[\protect\citeauthoryear{Pel{\'a}ez, Mat{\'e}, Tanarro, Molpeceres,
  Jim{\'e}nez-Redondo, Tim{\'o}n, Escribano  \& Herrero}{Pel{\'a}ez
  et~al.}{2018}]{Pelaez2018-dl}
Pel{\'a}ez R.~J.,  Mat{\'e} B.,  Tanarro I.,  Molpeceres G.,
  Jim{\'e}nez-Redondo M.,  Tim{\'o}n V.,  Escribano R.,   Herrero V.~J.,  2018,
  \mn@doi [Plasma Sources Sci. Technol.] {10.1088/1361-6595/aab185}, 27, 035007

\bibitem[\protect\citeauthoryear{Pendleton \& Allamandola}{Pendleton \&
  Allamandola}{2002}]{Pendleton2002-tf}
Pendleton Y.~J.,  Allamandola L.~J.,  2002, \mn@doi [Astrophys. J. Suppl. Ser.]
  {10.1086/322999}, 138, 75

\bibitem[\protect\citeauthoryear{Pino et~al.,}{Pino et~al.}{2008}]{Pino2008-al}
Pino T.,  et~al., 2008, \mn@doi [Astron. Astrophys. Suppl. Ser.]
  {10.1051/0004-6361:200809927}, 490, 665

\bibitem[\protect\citeauthoryear{Reynaud, Guillois, Herlin-Boime, Rouzaud,
  Galvez, Clinard, Balanzat  \& Ramillon}{Reynaud
  et~al.}{2001}]{Reynaud2001-uy}
Reynaud C.,  Guillois O.,  Herlin-Boime N.,  Rouzaud J.~N.,  Galvez A.,
  Clinard C.,  Balanzat E.,   Ramillon J.~M.,  2001, \mn@doi [Spectrochim. Acta
  A Mol. Biomol. Spectrosc.] {10.1016/s1386-1425(00)00445-5}, 57, 797

\bibitem[\protect\citeauthoryear{Ristein, Stief, Ley  \& Beyer}{Ristein
  et~al.}{1998}]{Ristein1998-tu}
Ristein J.,  Stief R.~T.,  Ley L.,   Beyer W.,  1998, \mn@doi [J. Appl. Phys.]
  {10.1063/1.368563}, 84, 3836

\bibitem[\protect\citeauthoryear{Sadezky, Muckenhuber, Grothe, Niessner  \&
  P{\"o}schl}{Sadezky et~al.}{2005}]{Sadezky2005-ii}
Sadezky A.,  Muckenhuber H.,  Grothe H.,  Niessner R.,   P{\"o}schl U.,  2005,
  \mn@doi [Carbon N. Y.] {10.1016/j.carbon.2005.02.018}, 43, 1731

\bibitem[\protect\citeauthoryear{Salama}{Salama}{2019}]{Salama2019-ms}
Salama F.,  2019, \mn@doi [Proc. Int. Astron. Union]
  {10.1017/S1743921320004123}, 15, 281

\bibitem[\protect\citeauthoryear{Sandford et~al.,}{Sandford
  et~al.}{2006}]{Sandford2006-gk}
Sandford S.~A.,  et~al., 2006, \mn@doi [Science] {10.1126/science.1135841},
  314, 1720

\bibitem[\protect\citeauthoryear{Santoro et~al.,}{Santoro
  et~al.}{2020}]{Santoro2020-ta}
Santoro G.,  et~al., 2020, \mn@doi [Astrophys. J.] {10.3847/1538-4357/ab9086},
  895

\bibitem[\protect\citeauthoryear{Schultrich}{Schultrich}{2018}]{Schultrich2018-qc}
Schultrich B.,  2018, in Springer series in materials science, Tetrahedrally
  Bonded Amorphous Carbon Films {I}.
Springer Berlin Heidelberg, Berlin, Heidelberg, pp 111--192,
  \mn@doi{10.1007/978-3-662-55927-7_5}

\bibitem[\protect\citeauthoryear{Sciamma-O'Brien \& Salama}{Sciamma-O'Brien \&
  Salama}{2020}]{Sciamma-OBrien2020-bm}
Sciamma-O'Brien E.,  Salama F.,  2020, \mn@doi [ApJ]
  {10.3847/1538-4357/abc00d}, 905, 45

\bibitem[\protect\citeauthoryear{Scott \& Duley}{Scott \&
  Duley}{1996}]{Scott1996-op}
Scott A.,  Duley W.~W.,  1996, \mn@doi [Astrophys. J. Lett.] {10.1086/310365},
  472

\bibitem[\protect\citeauthoryear{Shinohara, Cho, Shibata, Okamoto, Nakatani,
  Matsuda  \& Fujiyama}{Shinohara et~al.}{2008}]{Shinohara2008-bq}
Shinohara M.,  Cho K.,  Shibata H.,  Okamoto K.,  Nakatani T.,  Matsuda Y.,
  Fujiyama H.,  2008, \mn@doi [Thin Solid Films] {10.1016/j.tsf.2007.10.014},
  516, 4379

\bibitem[\protect\citeauthoryear{Shinohara, Tominaga, Shimomura, Ihara, Yagyu,
  Ohshima  \& Kawasaki}{Shinohara et~al.}{2018}]{Shinohara2018-io}
Shinohara M.,  Tominaga T.,  Shimomura H.,  Ihara T.,  Yagyu Y.,  Ohshima T.,
  Kawasaki H.,  2018, \mn@doi [IEEJ Trans. Fundam. Mater.]
  {10.1541/ieejfms.138.544}, 138, 544

\bibitem[\protect\citeauthoryear{Sleno}{Sleno}{2012}]{Sleno2012-ri}
Sleno L.,  2012, \mn@doi [J. Mass Spectrom.] {10.1002/jms.2978}, 47

\bibitem[\protect\citeauthoryear{Tielens}{Tielens}{2022}]{Tielens2022-bp}
Tielens A. G. G.~M.,  2022, \mn@doi [Front. Astron. Space Sci.]
  {10.3389/fspas.2022.908217}, 9

\bibitem[\protect\citeauthoryear{Tuinstra \& Koenig}{Tuinstra \&
  Koenig}{1970}]{Tuinstra1970-jg}
Tuinstra F.,  Koenig J.~L.,  1970, \mn@doi [J. Chem. Phys.]
  {10.1063/1.1674108}, 53, 1126

\bibitem[\protect\citeauthoryear{Wang, Alsmeyer  \& McCreery}{Wang
  et~al.}{1990}]{Wang1990-xp}
Wang Y.,  Alsmeyer D.~C.,   McCreery R.~L.,  1990, \mn@doi [Chem. Mater.]
  {10.1021/cm00011a018}, 2, 557

\bibitem[\protect\citeauthoryear{Wright, Bridger, Geballe  \& Pendleton}{Wright
  et~al.}{1996}]{Wright1996-dx}
Wright G.~S.,  Bridger A.,  Geballe T.~R.,   Pendleton T.,  1996, New
  Extragalactic Perspectives in the New South Africa.
Kluwer, Dordrecht, \mn@doi{10.1007/978-94-009-0335-7_16}

\bibitem[\protect\citeauthoryear{Ziegler, Ziegler  \& Biersack}{Ziegler
  et~al.}{2010}]{Ziegler2010-un}
Ziegler J.~F.,  Ziegler M.~D.,   Biersack J.~P.,  2010, \mn@doi [Nucl. Instrum.
  Methods Phys. Res. B] {10.1016/j.nimb.2010.02.091}, 268, 1818

\bibitem[\protect\citeauthoryear{van Dishoeck}{van
  Dishoeck}{2019}]{Van_Dishoeck2019-ui}
van Dishoeck E.~F.,  2019, \mn@doi [Proc. Int. Astron. Union]
  {10.1017/S1743921319008792}, 15, 3

\makeatother
\end{thebibliography}

%\begin{thebibliography}{99}
%\end{thebibliography}

%%%%%%%%%%%%%%%%%%%%%%%%%%%%%%%%%%%%%%%%%%%%%%%%%%

%%%%%%%%%%%%%%%%% APPENDICES %%%%%%%%%%%%%%%%%%%%%

%\appendix


%If you want to present additional material which would interrupt the flow of the main paper, it can be placed in an Appendix which appears after the list of references.

%%%%%%%%%%%%%%%%%%%%%%%%%%%%%%%%%%%%%%%%%%%%%%%%%%

% Don't change these lines
\bsp	% typesetting comment
\label{lastpage}
\end{document}